\documentclass[pre,aps,twocolumn,showpacs,amsmath,amssymb,amsfonts,floatfix,superscriptaddress]{revtex4}

\usepackage{graphicx}% include figure files
\usepackage{dcolumn}% align table columns on decimal point
\usepackage{bm}% bold math
\usepackage{hyperref}
\usepackage{latexsym}
\usepackage{float}
\usepackage{supertabular}
\usepackage{longtable}
\usepackage{amsmath}
\usepackage{amssymb}
\usepackage{bm}

\begin{document}
\title{Characterization of spiraling patterns in spatial rock--paper--scissors games}

\author{Bartosz Szczesny}
\email{mmbs@leeds.ac.uk}
\affiliation{Department of Applied Mathematics, School of Mathematics, University of Leeds, Leeds LS2 9JT, U.K.}

\author{Mauro Mobilia}
\email{M.Mobilia@leeds.ac.uk}
\affiliation{Department of Applied Mathematics, School of Mathematics, University of Leeds, Leeds LS2 9JT, U.K.}

\author{Alastair M. Rucklidge}
\email{A.M.Rucklidge@leeds.ac.uk}
\affiliation{Department of Applied Mathematics, School of Mathematics, University of Leeds, Leeds LS2 9JT, U.K.}

\begin{abstract}

The spatio--temporal arrangement of interacting populations often influences the maintenance of species diversity and is a subject of intense research. Here, we study the spatio--temporal patterns arising from the cyclic competition between three species in two dimensions. Inspired by recent experiments, we consider a generic metapopulation model comprising ``rock--paper--scissors'' interactions via dominance removal and replacement, reproduction, mutations, pair--exchange and hopping of individuals. By combining analytical and numerical methods, we obtain the model's phase diagram near its Hopf bifurcation and quantitatively characterize the properties of the spiraling patterns arising in each phase. The phases characterizing the cyclic competition away far from the Hopf bifurcation (at low mutation rate) are also investigated. Our analytical approach relies on the careful analysis of the properties of the complex Ginzburg--Landau equation derived through a controlled (perturbative) multiscale expansion around the model's Hopf bifurcation. Our results allows us to clarify when spatial ``rock--paper--scissors'' competition leads to stable spiral waves and under which circumstances they are influenced by nonlinear mobility.

\end{abstract}
\maketitle

\section{Introduction}

Ecosystems consist of a large number of interacting organisms and species organized in rich and complex evolving structures~\cite{May,patterns}. The understanding of what helps maintain biodiversity is of paramount importance for the characterization of ecological and biological systems. In this context, local interactions and the spatial arrangement of individuals have been found to be closely related to the stability and coexistence of species, and has therefore received significant attention~\cite{Huntley}. In particular, cyclic dominance has been shown to be a motif facilitating the coexistence of diverse species in a number of ecosystems ranging from side--blotched lizards~\cite{Sinervo,LizardMutation} and communities of bacteria~\cite{Kerr,Kirkup,Nahum} to plants systems and coral reef invertebrates~\cite{plants,Jackson}. It is noteworthy that cyclic dominance is not restricted only to biological systems but has also been found in models of behavioral science~\cite{OnNetworks}, e.g. in some public goods games~\cite{public-goods}. Remarkably, experiments on three strains of {\it E.coli} bacteria in cyclic competition on two--dimensional plates yield spatial arrangements that were shown to sustain the long--term coexistence of the species~\cite{Kerr}. Cyclic competitions of this type have been modeled with rock--paper--scissors (RPS) games, where ``rock crushes scissors, scissors cut paper, and paper wraps rock''~\cite{games}.

While non--spatial RPS--like games usually drive all species but one to extinction in finite time~\cite{RPSnonspatial}, their spatial counterparts are generally characterized by intriguing complex spatio--temporal patterns sustaining the species coexistence, see e.g. Refs.~\cite{zerosum,RMF,RF,Matti,He,SMR,FigshareMovies}. In recent years, many models for the RPS cyclic competition have been considered. In particular, various two--dimensional versions of the model introduced by May and Leonard~\cite{MayLeonard} have been studied~\cite{RMF,RF,Matti,He,SMR,rulands13}. In spatial variants of the May--Leonard model, it was found that mobility implemented by pair--exchange among neighbors can significantly influence species diversity: below a certain mobility threshold species coexist over long periods of time and self--organize by forming fascinating spiraling patterns, whereas biodiversity is lost when that threshold is exceeded~\cite{RMF}. Other popular RPS models are those characterized by a conservation law at mean field level (``zero--sum'' games). In two spatial dimensions, these zero--sum models are also characterized by a long--lasting coexistence of the species, but in this case the population does not form spiraling patterns~\cite{zerosum}. On the other hand, while microbial communities in cyclic competition were found to self--organize in a complex manner, it is not clear whether there is a parameter regime in which their spatial arrangement would form spirals as those observed in myxobacteria and in Dictyostelium mounds~\cite{spirals_exp}. In this context, we believe that this work contributes to understanding the relationship between the maintenance of species diversity and the formation of spiraling patterns in populations in cyclic competitions.

To shed further light on the evolution and self--organization of population in cyclic competition, in this work, we comprehensively characterize the spatio--temporal properties of a generic two--dimensional model for the cyclic competition between three species that unifies the various processes considered in Refs.~\cite{RMF,RF,Matti,He,rulands13}. The model that we consider accounts for cyclic competition with {\it dominance--removal}~\cite{RMF,Matti,He,rulands13} and {\it dominance--replacement}~\cite{zerosum}, also including reproduction, mutation and mobility in the form of hopping and pair--exchange between nearest neighbors. Our approach is inspired by the experiments of~\cite{Nahum} and the model is formulated at the metapopulation level~\cite{metapopulation,TuringNoisy}, which allows us to establish a close relationship between the underlying stochastic and deterministic dynamics. Within such a framework, we combine analytical and numerical methods to carefully analyze the properties of the emerging spatio--temporal patterns. Our main analytical tool consists of deriving a {\it complex Ginzburg--Landau} equation (CGLE)~\cite{CGLE} using a multiscale perturbative expansion in the vicinity of the model's Hopf bifurcation. The CGLE allows us to accurately analyze the spatio--temporal dynamics in the vicinity of the bifurcation and to faithfully describe the quantitative properties of the spiraling patterns arising in the four phases reported in Ref.~\cite{SMR,FigshareMovies}. Our theoretical predictions are fully confirmed by extensive computer simulations at different levels of description. We also study the system's phase diagram far from the Hopf bifurcation, where it is characterized by three phases, and show that the properties of the spiraling patterns can still be inferred from the CGLE. For this, we study phenomena like far--field break--up and convective instability of spiral waves discussing, and discuss how these are influenced by nonlinear mobility and enhanced cyclic dominance. 

Our paper is structured as follows: In Sec.~II, the generic metapopulation model~\cite{metapopulation} is introduced and its mean field analysis is presented. We also present the spatial deterministic description of the model with {\it nonlinear diffusion} and the perturbative derivation of the  CGLE. Section II is complemented by two technical appendices. The model's phase diagram near the Hopf bifurcation is studied in detail in Sec.~III where the CGLE is employed to characterize the properties of spiraling patterns in each phase. Section IV is dedicated to the analysis of the phase diagram, and to the properties of the spiraling patterns, far from the Hopf bifurcation and addresses how these are influenced by nonlinear mobility and by enhancing the rate of cyclic dominance. Finally, we conclude with a discussion and interpretation of our findings.

\section{The metapopulation model}

Spatial rock--paper--scissors games have mostly been studied on square lattices whose nodes can be either empty or at most occupied by one individual with the dynamics implemented via nearest--neighbor interactions~\cite{zerosum,RMF,RF,Matti,He}. Here, inspired by the experiments of Ref.~\cite{Nahum}, as well as by the works~\cite{Kerr,LizardMutation}, we adopt an alternative modeling approach in terms of a metapopulation model that allows further analytical progress. 

\begin{figure}
	\includegraphics[width=0.85\linewidth]{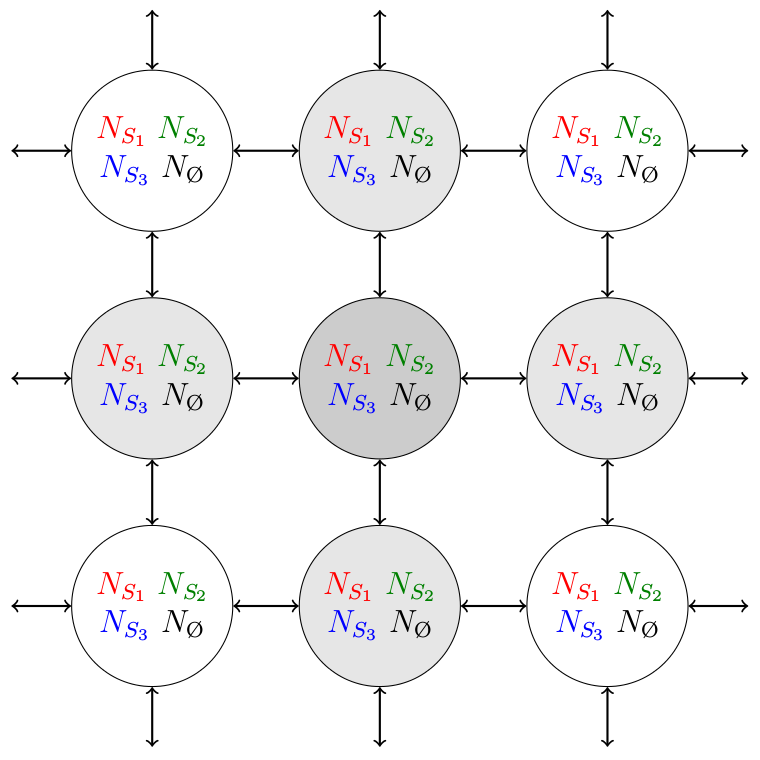}
	\caption{(Color online). Cartoon of the metapopulation model: $L \times L$ patches (or islands) are arranged on a periodic square lattice (of linear size $L$). Each patch ${\bm \ell}=(\ell_1, \ell_2)$ can accommodate at most $N$ individuals of species $S_1, S_2$, $S_3$ and empty spaces denoted $\O$. Each patch consists of a well--mixed population of $N_{S_1}$ individuals of species $S_1$, $N_{S_2}$ of type $S_2$, $N_{S_3}$ of type $S_3$ and $N_{\O} = N - N_{S_1} - N_{S_2} - N_{S_3}$ empty spaces. The composition of a patch evolves in time according to the processes (\ref{sel}) and (\ref{non-sel}). Furthermore, migration from the focal patch (dark gray) to its four nearest--neighbor (light gray) occurs according to the processes (\ref{migr}), see text.}
	\label{metapop}
\end{figure}

In the metapopulation formulation~\cite{SMR,FigshareMovies}, the lattice consists of a periodic square array of $L \times L$ patches (or islands) each of which comprises a well--mixed sub--population of constant size $N$ (playing the role of the carrying capacity) consisting of individuals of three species, $S_1$, $S_2$, $S_3$ and empty spaces ($\O$). It has to be noted that slightly different metapopulation models of similar systems have been recently considered, see e.g.~\cite{rulands13,lamouroux12}. As sketched in Fig.~\ref{metapop}, each patch of the array is labeled by a vector ${\bm \ell}=(\ell_1,\ell_2)$, with $\ell_{1,2}\in \{1,2,\dots, L\}$ and periodic boundary conditions, and can accommodate at most $N$ individuals, i.e. all patches have a carrying capacity $N$. Each patch ${\bm \ell}$ consists of a well--mixed (spatially unstructured) population comprising $N_{i}({\bm \ell})$ individuals of species $S_i$ ($i=1,2,3$) and $N_{\O}({\bm \ell})=N-N_{S_1}({\bm \ell})-N_{S_2}({\bm \ell})-N_{S_3}({\bm \ell})$ empty spaces.  Species $S_1$, $S_2$ and $S_3$ are in cyclic competition within each patch ({\it intra--patch interaction}), while all individuals can move to neighboring sites ({\it inter--patch mobility}), see below.

The population dynamics is implemented by considering the most generic form of cyclic rock--papers--scissors--like competition between the three species with the population composition {\it within each patch} evolving according to the following schematic reactions:
	\begin{eqnarray}
	\label{sel}
		S_i + S_{i+1} \xrightarrow{\sigma} S_i + \O &~&
		S_i + S_{i+1} \xrightarrow{\zeta } 2 S_i \\
	\label{non-sel}
		S_i + \O \xrightarrow{\beta} 2 S_i &~&
		S_i \xrightarrow{\mu} S_{i\pm1},
	\end{eqnarray}
where the species index $i \in \{1,2,3\}$ is \textit{ordered cyclically} such that $S_{3+1} \equiv S_1$ and $S_{1-1} \equiv S_3$. The reactions (\ref{sel}) describe the generic form of cyclic competition where $S_i$ dominates over $S_{i+1}$ and is dominated by $S_{i-1}$. They account for the {\it dominance--removal} selection processes (with rate $\sigma$) of Refs.~\cite{RMF,He}, as well as the {\it dominance--replacement} (zero--sum) processes (with rate $\zeta$) studied notably in \cite{zerosum}. The processes (\ref{non-sel}) allow for the reproduction of each species (with rate $\beta$) independently of the cyclic interaction provided that free space ($\O$) is available within the patch. Mutations of the type $S_i \xrightarrow{} S_{i\pm1}$ (with rate $\mu$) capture the fact that {\it E.~coli} bacteria are known to mutate~\cite{Kerr}, while the side--blotched lizards {\it Uta stansburiana} have been found to undergo throat--colour transformations~\cite{LizardMutation}. From a modeling viewpoint, the mutation yields a bifurcation around which considerable mathematical progress is feasible, see Sec.~\ref{nearHB} and \cite{SMR}.

\subsection{Mean field analysis}

When $N \to \infty$, demographic fluctuations are negligible and the population composition within each single patch is described by the continuous variables $s_i = N_{i} / N$ which obey the mean field rate equations (REs) derived in Appendix A
	\begin{eqnarray}
	\label{RE}
		\frac{d s_i}{d t}
		&=& s_i [\beta (1 - r) - \sigma s_{i-1} + \zeta (s_{i+1} - s_{i-1})] \nonumber \\ 
		&~& + \mu (s_{i-1} + s_{i+1} - 2 s_i) ,
	\end{eqnarray}
where ${\bm s}\equiv (s_1, s_2, s_3)$ and $r \equiv s_1 + s_2 + s_3$ is the total density and, since the carrying capacity is fixed, we have used $N_{\O} / N = 1 - r$. The REs~(\ref{RE}) admit a coexistence fixed point ${\bm s}^*=s^* (1,1,1)$ with $s^*=\beta/(3\beta + \sigma)$ that, in the presence of a non--vanishing mutation rate, is an asymptotically stable focus when $\mu > \mu_H = \frac{\beta\sigma}{6(3\beta + \sigma)}$ and is unstable otherwise. In fact, the REs~(\ref{RE}) are characterized by a supercritical Hopf bifurcation (HB) yielding a stable limit cycle of frequency close to $\omega_H = \frac{\sqrt{3}\beta (\sigma + 2\zeta)}{2(3\beta + \sigma)}$ when $\mu<\mu_H$~\cite{SMR} (see also Ref.~\cite{RPSmut}). In the absence of mutations ($\mu=0$), the coexistence state ${\bm s}^*$ is never asymptotically stable and the REs (\ref{RE}) yield either heteroclinic cycles (when $\mu=0$ and $\sigma>0$)~\cite{MayLeonard} or neutrally stable periodic orbits (when $\mu=\sigma=0$)~\cite{games}. In the absence of spatial structure, finite--size fluctuations are responsible for the rapid extinction of two species in each of these two cases~\cite{RPSnonspatial}. It is worth noting that the heteroclinic cycles are degenerate when $\sigma>0$ and $\zeta=\mu=0$.

\subsection{Dynamics with partial differential equations}

Since we are interested in analyzing the spatio--temporal arrangement of the populations, in addition to the intra--patch reactions (\ref{sel})-(\ref{non-sel}), we also allow individuals to migrate between neighboring patches ${\bm \ell}$ and ${\bm \ell}'$, according to 
	\begin{eqnarray}
	\label{migr}
		\big[S_i \big]_{{\bm \ell}} \big[\O \big]_{{\bm \ell}'} &\xrightarrow{\delta_D}&
		\big[\O	\big]_{{\bm \ell}} \big[S_i \big]_{{\bm \ell}'} \quad, \nonumber \\
		\big[S_i \big]_{{\bm \ell}} \big[S_{i\pm1} \big]_{{\bm \ell}'} &\xrightarrow{\delta_E}&
		\big[S_{i\pm1} \big]_{{\bm \ell}} \big[S_i \big]_{{\bm \ell}'},
	\end{eqnarray}
where pair--exchange (with rate $\delta_E$) is divorced from hopping (with rate $\delta_D$). In biology, organisms are in fact known not to simply move diffusively, but to sense and respond to their environment, see e.g. \cite{Kearns}. Here, (\ref{migr}) allows us to discriminate between the movement in crowded regions, where mobility is dominated by pair--exchange, and mobility in diluted regions where hopping can be more efficient, and leads to {\it nonlinear mobility} when $\delta_E \neq \delta_D$, see below and Refs.~\cite{SMR,Fanelli}.

The metapopulation formulation of the model defined by (\ref{sel})-(\ref{non-sel}) and (\ref{migr}) is ideally suited for a size expansion in the inverse of the carrying capacity $N$ of the underlying Master equation~\cite{SizeExp}. As shown in Appendix~\ref{AppendixA}, in the continuum limit and to lowest order, the master equation yields the following partial differential equations (PDEs) with periodic boundary conditions
	\begin{eqnarray}
	\label{pde}
		\partial_t s_i
		&=& s_i[\beta (1-r) - \sigma s_{i-1}] \nonumber\\
		&+& \zeta s_i[s_{i+1} - s_{i-1}]\nonumber\\ 
		&+& \mu \left[s_{i-1} + s_{i+1} - 2 s_i\right] \nonumber \\ 
				&+& (\delta_E-\delta_D) \left[r \Delta s_i - s_i
		\Delta r \right]\nonumber\\
		&+&\delta_D \Delta s_i,
	\end{eqnarray}
where here $s_i\equiv s_i({\bm x},t)$ and the contribution proportional to $\delta_E-\delta_D$ is a nonlinear diffusive term.
These PDEs give the continuum description of the system's deterministic dynamics on a domain of fixed size ${\cal S} \times {\cal S}$ defined on a square lattice comprising $L \times L$ sites with periodic boundary conditions, when $L \to \infty$ and ${\bm x}= {\cal S}({\bm \ell}/L)$ such that ${\bm x}\in [0, {\cal S}]^2$. In such a setting, the mobility rates of (\ref{migr}) are rescaled according to $\delta_{D,E} \to \delta_{D,E} (\frac{{\cal S}}{L})^2$ and interpreted as diffusion coefficients (see Appendix~\ref{AppendixA}). However, to mirror the properties of the metapopulation lattice model, throughout this paper we use ${\cal S}=L$. We have found that the choice ${\cal S}=L$ is well--suited to describe spatio--temporal patterns whose size exceeds the unit spacing, as is always the case in this work. Eqs (\ref{pde}) have been solved using the second order exponential time differencing method (with a time step $\delta t$ = 0.125) with fast Fourier transforms with a number of modes ranging from $128\times 128$ to $8192\times 8192$~\cite{CoxMatthews,FFT}.
 
Even though the derivation of (\ref{pde}) assumes $N \gg 1$ (see Appendix~\ref{AppendixA}), as illustrated in Fig.~\ref{LatticevsPDE} (see also \cite{SMR,FigshareMovies}), it has been found that (\ref{pde}) accurately capture the properties of the lattice model, whose dynamics is characterized by the emergence of fascinating spiraling patterns, when $N \gtrsim 20$ and $\mu<\mu_H$ (no coherent patterns are observed when $\mu>\mu_H$)~\cite{FigshareMovies}. When $N=4-16$ the outcomes of stochastic simulations are noisy but, quite remarkably, it also turns out that the solutions of (\ref{pde}) still reproduce some of the outcomes of stochastic simulations~\cite{SMR,FigshareMovies}, see Sec.~\ref{awayHB}. In Figure~\ref{LatticevsPDE}, as in all the other figures, the results of stochastic and deterministic simulations are visualized by color coding the abundances of the three species in each patch with appropriate RGB intensities such that (red, green, blue) $= (s_1, s_2, s_3)$ resulting in empty spaces being color--coded in black.

\begin{figure}
	\includegraphics[width=\linewidth]{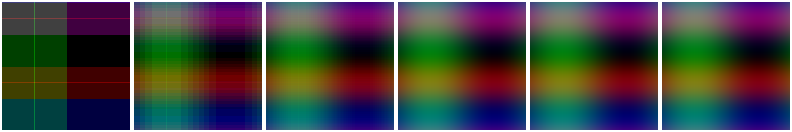}
	\includegraphics[width=\linewidth]{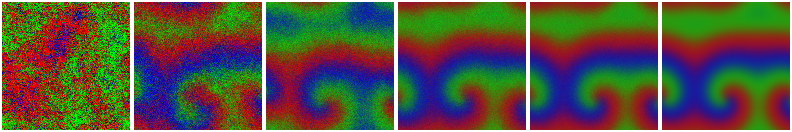}	
	\caption{(Color online). Comparison of lattice simulations (performed using a spatial Gillespie algorithm~\cite{Gillespie}) with solutions of (\ref{pde}) in the bound state phase (BS), where the spiral waves are stable, near the HB point, see text of Sec.~\ref{nearHB}. Rightmost panels show the solutions of (\ref{pde}) while the remaining panels show results of stochastic simulations for $L^2 = 128^2$ with $N = 4, 16, 64, 256, 1024$ (from left to right). As in all other figures, each color represents one species with dark dots indicating low density regions. Top panels show initial conditions while the lower panels show the domains at $t = 1000$. The other parameters are $\beta=\sigma=\delta_D=\delta_E=1,\zeta=0.6$ and $\mu=0.02$.}
	\label{LatticevsPDE}
\end{figure}

To next--to--leading order, the size expansion of the master equation yields a Fokker--Planck equation that can be used for instance to characterize the system's spatio--temporal properties in terms of its power spectra, see e.g.~\cite{TuringNoisy,RPSnonspatial,RPSmut}. Here, we adopt a different route and will show that the emerging spiraling patterns can be comprehensively characterized from the properties of a suitable CGLE properly derived from (\ref{pde}).

\subsection{Complex Ginzburg--Landau equation}
\label{cgle}

The complex Ginzburg--Landau equation (CGLE) is well--known for its rich phase diagram characterized by the formation of complicated coherent structures, like spiral waves in two dimensions, see e.g.~\cite{CGLE}.

In the context of spatial RPS games, the properties of the CGLE have been used first in Refs.~\cite{RMF} for a variant of the model considered here with only dominance--removal competition ($\zeta=\mu=0$ and $\delta_D=\delta_E$). The treatment was then extended to also include dominance--replacement competition (with $\mu=0$ and $\delta_D=\delta_E$)~\cite{RF,rulands13}, and has  recently been generalized to more than three species~\cite{pleimling14}. In all these works, the derivation of the CGLE relies on the fact that the underlying mean field dynamics quickly settles on a two--dimensional manifold on which the flows approach the absorbing boundaries forming heteroclinic cycles~\cite{MayLeonard,games}. These are then treated as stable limit cycles and the spatial degrees of freedom are reinstated by introducing linear diffusion (see also \cite{Orihashi}). While this approach remarkably succeeded in explaining various properties of the underlying models upon adjusting (fitting) one parameter, it rests on a number of uncontrolled steps. These include the approximation of heteroclinic cycles by stable limit cycles and the omission of the nonlinear diffusive terms that arise from the transformations leading to the CGLE~\cite{games}.

Here, we consider an alternative derivation of the CGLE that approximates (\ref{pde}) and describes the properties of the generic metapopulation model defined by (\ref{sel})-(\ref{non-sel}) and (\ref{migr}). Since the mean field dynamics is characterized by a stable limit cycle (when $\mu<\mu_H$) resulting from a Hopf bifurcation (HB) arising at $\mu=\mu_H$, our approach builds on a perturbative multiscale expansion around $\mu_H$ (HB point). For this, we proceed with a space and time perturbation expansion in the parameter $\epsilon= \sqrt{3(\mu_H - \mu)}$ ~\cite{SMR} in terms of the ``slow variables'' $({\bm X}, T)=(\epsilon{\bm x},~\epsilon^2 t)$~\cite{Miller,Manneville}. While the details of the derivation are provided in Appendix~\ref{AppendixB}, we here summarize the main steps of the analysis. After the transformation ${\bm s}\to {\bm u}={\bm M}(\bm{s}- \bm{s}^*)$, where ${\bm u}=(u_1,u_2,u_3)$ and ${\bm M}$ is given by (\ref{M}), $u_3$ decouples from $u_1$ and $u_2$ (to linear order), and one writes $\bm{u}(\bm{x},t) = \sum_{n=1}^{3} \epsilon^n \bm{U}^{(n)}(t,T,\bm{X})$, where the components of $\bm{U}^{(n)}$ are of order ${\cal O}(1)$. Substituting into (\ref{pde}), with $U_1^{(1)}+iU_2^{(1)}=\mathcal{A}(T,\bm{X})e^{i\omega_H t}$, one finds that $\mathcal{A}$ is a modulated complex amplitude satisfying a CGLE obtained by imposing the removal of the secular term arising at order $\mathcal{O}(\epsilon^3)$, see Appendix~\ref{AppendixB} and Ref.~\cite{SMR}.  Upon rescaling $\mathcal{A}$ by a constant (see Appendix~\ref{AppendixB}), this yields the two--dimensional CGLE with a real diffusion coefficient $\delta= \frac{3\beta\delta_E + \sigma\delta_D}{3\beta + \sigma}$:
	\begin{equation}
	\label{CGLE}
		\partial_{T} \mathcal{A} =
		\delta \Delta_{\bm X} \mathcal{A} + \mathcal{A} - (1 + i c) |\mathcal{A}|^2 \mathcal{A},
	\end{equation}
where  $\Delta_{\bm X}= \partial_{X_1}^2+\partial_{X_2}^2=\epsilon^{-2}(\partial_{x_1}^2+\partial_{x_2}^2)$ and
	\begin{equation}
	\label{c3}
		c = \frac
		{12\zeta (6\beta - \sigma)(\sigma + \zeta) + \sigma^2 (24\beta - \sigma)}
		{3\sqrt{3} \sigma (6\beta + \sigma)(\sigma + 2\zeta)}.
	\end{equation}

At this point it is worth noting the following:
	\begin{enumerate}
		\item[(i)] The CGLE (\ref{CGLE}) is a controlled approximation of the the PDEs (\ref{pde}) around the HB and its expression differs from those obtained in a series of previous works, e.g, in ~\cite{RMF,Matti,RF,rulands13,pleimling14}. In particular, the functional dependence of the CGLE parameter (\ref{c3}) differs from that used in Refs.~\cite{RMF,Matti,RF,rulands13,pleimling14} for the special cases $\mu=\zeta=0$ and $\mu=0$.
		\item[(ii)] As shown in Section~\ref{nearHB}, the phase diagram and the emerging spiraling patterns around the HB can be quantitatively described in terms of the sole parameter $c$, given by (\ref{c3}), that does not depend on $\mu$ (since here $\mu\approx \mu_H$). 
		\item[(iii)] It has to be stressed that in the derivation of (\ref{CGLE}) no nonlinear diffusive terms appear at order $\mathcal{O}(\epsilon^3)$. In fact, the perturbative multiscale expansion yields the CGLE (\ref{CGLE}) with only a linear diffusion term $\delta \Delta_{\bm X} \mathcal{A}$, where $\delta=\delta(\delta_D,\delta_E)$ is an effective diffusion coefficient that reduces to $\delta_E$ when $\beta \gg \sigma$ and to $\delta \to \delta_D$ when $\beta\ll \sigma$~\cite{SMR}. This implies that nonlinear mobility plays no relevant role near the HB where mobility merely affects the spatial scale but neither the system's phase diagram nor the stability of the ensuing patterns. Near the HB, one can therefore set $\delta_E=\delta_D=1$ yielding $\delta=1$ without loss of generality.
	\end{enumerate}

\begin{figure}
	\includegraphics[width=\linewidth]{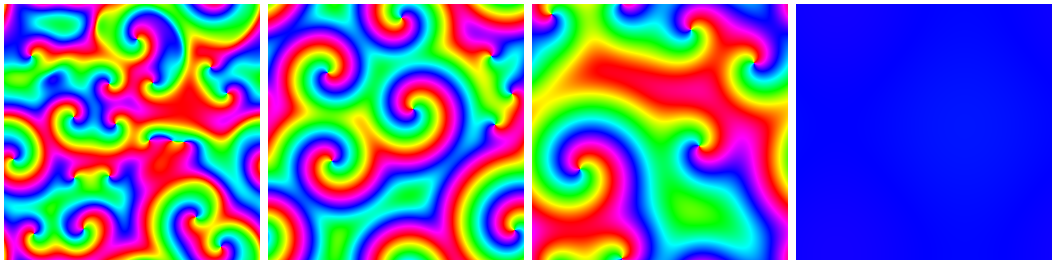}
	\caption{(Color online). Four phases in the two--dimensional CGLE (\ref{CGLE}) for $c = (2.0, 1.5, 1.0, 0.5)$ from left to right. Spiral waves of the third panel (from the left) are stable while the others are unstable, see Sec.~\ref{nearHB}. Here, the colors represent the argument of $\mathcal{A}$ encoded in hue: red, green and blue respectively correspond to arguments $0$, $\pi/3$ and $2\pi/3$.}
	\label{CGLEarg}
\end{figure}

In Sections~\ref{nearHB} and \ref{awayHB}, we show how the properties of the CGLE (\ref{CGLE}) can be used to obtain the system's phase diagram and to comprehensively characterize the oscillating patterns emerging in four different phases around the HB, and also to gain significant insight into the system's spatio--temporal behavior away from the HB. 
For the sake of simplicity we here restrict $\sigma$ and $\zeta$ into $[0,4]$. Since the components of ${\bm u}={\bm M}({\bm s}-{\bm s}^*)$ are linear superposition of the species' densities and $\mathcal{A}({\bm X},T)=e^{-i\omega_H t}(U_1^{(1)} + i U_2^{(1)})$, the modulus $|\mathcal{A}|$ of the solution of (\ref{CGLE}) is bounded by $0$ and $1$ when one works with the slow $({\bm X},T)$--variables. Hence, as illustrated by Fig.~\ref{CGLEarg}, the argument of $\mathcal{A}$ carries useful information on the wavelength and speed of the patterns, whereas its modulus allows us to track the position of the spiral cores, identified as regions where $|\mathcal{A}|\approx 0$ corresponding to close to zero deviations from the steady state ${\bm s}^*$ (see Fig.~\ref{SA2} below).

\section{State diagram near the Hopf bifurcation \& characterization of four phases}
\label{nearHB}

\begin{figure}
	\includegraphics[width=\linewidth]{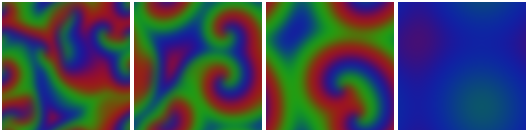}
	\includegraphics[width=\linewidth]{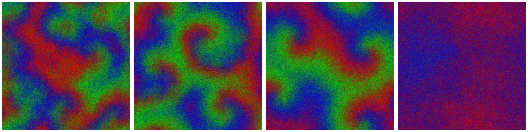}\\
	~\\
	\includegraphics[width=\linewidth]{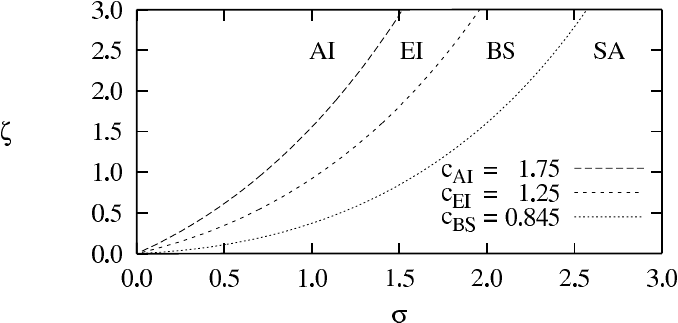}
	\caption{(Color online). Upper panels: Typical snapshots of the phases AI, EI, BS, SA (from left to right) as obtained from  (\ref{pde}) (top row) and from lattice simulations (middle row) with parameters $\sigma=\beta = \delta_E=\delta_D=1$, $\mu=0.02$, $L=128, N=64$ and, from left to right, $\zeta=(1.8, 1.2, 0.6, 0)$. The corresponding values of the CGLE parameter (\ref{c3}) are $c\approx (1.94, 1.47, 1.01, 0.63)$. Lower panel: Phase diagram of the two--dimensional RPS system around the Hopf bifurcation with contours of $c = (c_{{\rm AI}}, c_{{\rm EI}}, c_{{\rm BS}})$ in the $\sigma-\zeta$ plane, see text. We distinguish four phases: spiral waves are unstable in AI, EI and SA phases, while they are stable in BS phase. The boundaries between the phases have been obtained using (\ref{c3}), see Refs.~\cite{SMR,FigshareMovies} for details.}
 	\label{diagram_phases}
\end{figure}

The CGLE (\ref{CGLE}) enables us to obtain an accurate characterization of the spatio--temporal patterns in the vicinity of the HB by relying on the well--known phase diagram of the two--dimensional CGLE~\cite{CGLE}. The latter consists of four distinct phases which can be classified in terms of the CGLE parameter $c$ given by (\ref{c3})~\cite{SMR,FigshareMovies}. As illustrated in Fig.~\ref{diagram_phases}, these are separated by the three critical values $(c_{{\rm AI}}, c_{{\rm EI}}, c_{{\rm BS}}) \approx (1.75, 1.25, 0.845)$. In the {\it absolute instability (AI) phase}, arising when $c > c_{{\rm AI}}$, no stable spiral waves can be sustained. In the {\it Eckhaus instability (EI) phase}, arising when $c_{{\rm EI}} < c < c_{{\rm AI}}$, spiral waves are convectively unstable and their arms are first distorted and then break up. Spiral waves are stable in the {\it bound state (BS) phase} that arises when $c_{{\rm BS}} < c < c_{{\rm EI}}$. Spiral waves collide and annihilate in the {\it spiral annihilation (SA)} phase when $0 < c < c_{{\rm BS}}$.

As illustrated by Figs.~\ref{LatticevsPDE} and \ref{CGLEarg}, and in the upper panels of Fig.~\ref{diagram_phases}, we have verified for different sets of parameters $(\beta,\sigma,\zeta)$ and $c$ that the deterministic predictions of (\ref{pde}) and of the CGLE (\ref{CGLE}) correctly reflect the properties of the lattice metapopulation system, with a striking correspondence as soon as $N\gtrsim 64$.

In this section, Eq.~(\ref{CGLE}) is used to derive the system's phase diagram around the HB and to fully characterize each of its four phases. As explained below, the effect of noise has been found to significantly affect the dynamics only when the mobility rate is particularly low and $N$ is of order of the unity, see Sec.~\ref{awayHB}.B, but the spatio--temporal properties of the lattice model are well captured by (\ref{pde}) when the size of the patterns moderately exceeds that of lattice spacing, see Fig.~\ref{LatticevsPDE}. In what follows, our analysis is based mainly on (\ref{CGLE}) and we have carried out extensive numerical simulations confirming that (\ref{pde}) and the CGLE provide a faithful description of the lattice metapopulation model's dynamics when $N\gtrsim 16$, while their predictions have been found to also qualitatively reproduce some aspects of the lattice simulation when $N=2-16$, see \cite{SMR,FigshareMovies}.

\subsection{Bound state phase ($0.845 \lesssim c \lesssim 1.25$)}

\begin{figure}
	\includegraphics[width=\linewidth]{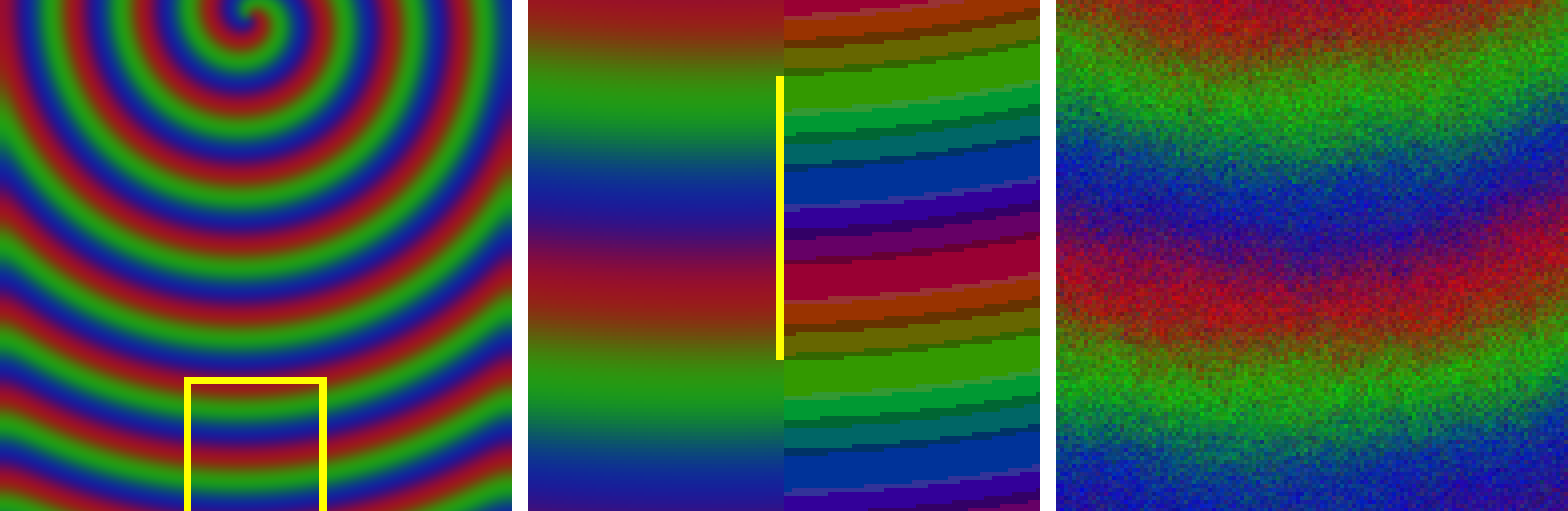}
	\caption{(Color online). Leftmost: Domain of size $512^2$ cut out from a numerical solution of  (\ref{pde}) with $\beta = \sigma = \delta_D = \delta_E = 1$, $\zeta = 0.3$, $\mu = 0.02$ and $L^2 = 1024^2$. The yellow frame outlines domain of size $128^2$ enlarged in the middle panel. Middle: Part of a spiral arm (far from the core) resembling a plane wave enlarged from the left panel. The color depth of the right half of the image was reduced to 256 colors for an easy identification of the wavelength  found to be equal to 71 length units in the physical domain as measured by the yellow bar. Rightmost: Same as in the middle panel from lattice simulations with $N=64$.}
	\label{BSphase}
\end{figure}

When $c_{{\rm BS}} < c < c_{{\rm EI}}$, the system lies in the bound state phase where the dynamics is characterized by the emergence of stable spiral waves that have a well--defined wavelength $\lambda$ and phase velocity $v$. This is fully confirmed by our lattice simulations and by the solutions of (\ref{pde}), as illustrated in Fig.~\ref{BSphase} where one observes well--formed spirals whose wavelengths are independent of $N$ and $L$. These quantities can be related analytically using the CGLE (\ref{CGLE}) by proposing a traveling plane wave ansatz $\mathcal{A}({\bm X},T)=R~e^{i({\bf k}.{\bm X}-\omega T)}$, where $R$ is the plane wave amplitude. Such a  traveling wave ansatz is a suitable approximation away from the core of the spiraling patterns as verified in our numerical simulations. Substitution into (\ref{CGLE}) gives $\omega=c R^2$ and $R^2=1-\delta k^2$ when the imaginary and real parts are equated, respectively. This yields the dispersion relation
	\begin{eqnarray}
	\label{disp-rel}
		\omega = c R^2 = c (1 - \delta k^2).
	\end{eqnarray}

This indicates that a plane wave is possible only when the wavenumber $k$ (modulus of the wave vector ${\bf k}$) satisfies $\delta k^2<1$.

\begin{figure}
	\includegraphics[width=\linewidth]{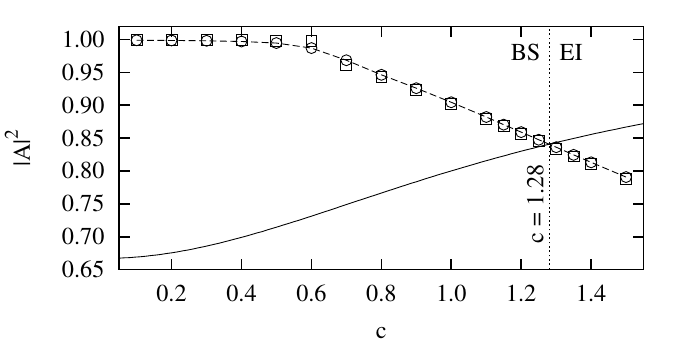}
	\caption{Numerical values of $|\mathcal{A}|^2$ obtained from a histogram with $1000$ bins (squares) and averaging (circles) with interpolation (dashed). When the traveling wave ansatz is valid (in BS and EI phases, away from the spirals' cores), $|\mathcal{A}|^2\approx R^2$, see text. Solid line is the theoretical Eckhaus criterion (\ref{EIpert}) obtained from the plane wave Ansatz yielding $c_{{\rm EI}} \approx 1.28$ marked by the dotted line. This has to be compared with the value of $c_{{\rm EI}} \approx 1.25$ reported in the phase diagram of the two--dimensional CGLE \cite{CGLE}. Spiral waves are convectively unstable in the region where $c > c_{{\rm EI}}$ and are stable just below that value in the BS phase, see Sec.~\ref{SecEI}.}
	\label{R2}
\end{figure}

\begin{figure}
	\includegraphics[width=0.9\linewidth]{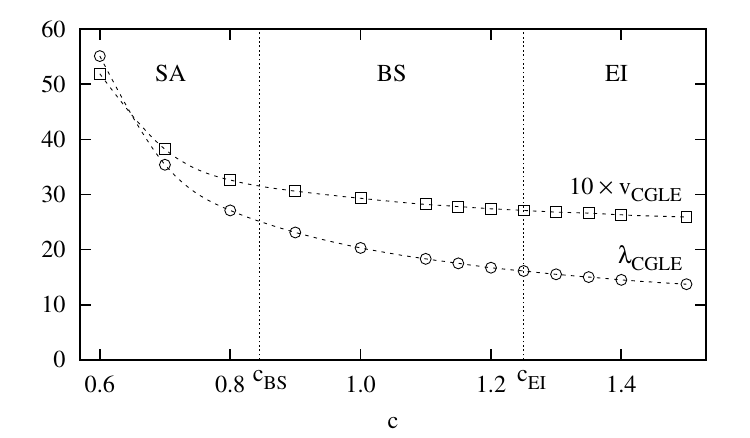}
	\caption{(Color online). Wavelength ($\circ$) and (rescaled) velocity ($\diamond$) obtained from the CGLE (\ref{CGLE}) with $\delta=1$ as functions of the parameter
 $c=0.6-1.5$. The critical values $c_{{\rm BS}}$ and $c_{{\rm EI}}$ separating the SA and BS phases and the BS and EI phases are indicated as thin dotted lines, see text.}
 	\label{lambda_vel}
\end{figure}

We have numerically found that $k$ and the wavelength of the spiraling patterns vary with the system parameters, as reported in Fig.~\ref{R2} where $|\mathcal{A}|^2$  is shown to decrease with $c$ in the range $0.845 \lesssim c \lesssim 1.25$, with $|\mathcal{A}|^2\approx R^2$ when the traveling wave ansatz is valid. The wavelength and phase velocity of the patterns can be obtained from the CGLE (\ref{CGLE}) and the dispersion relation (\ref{disp-rel}) by noting that $k=\sqrt{(1- R^2)/\delta}$ and therefore $\lambda_{{\rm CGLE}}=2\pi/k$ and $v_{{\rm CGLE}} = \omega/k$, see Fig.~\ref{lambda_vel}. At this point, it is important to realize that $\lambda_{{\rm CGLE}}$ and $v_{{\rm CGLE}}$ are expressed in terms of the slow $({\bm X},T)$-variables. By reinstating the physical units $({\bm x},t)=({\bm X}/\epsilon,T/\epsilon^2)$ one finds the spirals' physical wavelength
	\begin{eqnarray}
	\label{lambda}
		\lambda = \frac{ \lambda_{{\rm CGLE}}}{\epsilon} = \frac{2\pi}{\epsilon}~\sqrt{\frac{\delta}{1-R^2}}
	\end{eqnarray}
and velocity
	\begin{eqnarray}
	\label{vel}
		v = \epsilon~v_{{\rm CGLE}} = \epsilon~cR^2\sqrt{\frac{\delta}{1-R^2}}.
	\end{eqnarray}

Our numerical simulations have shown that both $k$ and the amplitude $R$ of the plane wave are nontrivial functions of the CGLE parameter $c$ given by (\ref{c3}), see Fig.~\ref{R2}. The theoretical predictions of the velocity and wavelength of the spiral waves have thus been obtained by substituting into (\ref{vel}) and (\ref{lambda}) the square of the plane wave amplitude $R^2$ by its value computed from the solutions of CGLE (with $\delta=1$) as a function of $c$, see Fig.~\ref{R2}. To this end, the numerical solutions of (\ref{CGLE}) have been integrated initially up to time $t = 799$ until the spirals are well developed to avoid any transient effects. Then, the amplitude from the successive $200$ data frames between $t = 800$ and $t = 999$ were averaged, yielding about $1.3~\times 10^7$ data points for each value of $c$. The results (for $\lambda_{{\rm CGLE}}$ and $v_{{\rm CGLE}}$) are summarized in Fig.~\ref{lambda_vel} which shows that the wavelength decreases monotonically when $c$ is increased (and $R$ decreases, see Fig.~\ref{R2}), with wavelengths ranging from $\lambda_{{\rm CGLE}} \approx 26$ to $\lambda_{{\rm CGLE}} \approx 16$ when $c$ varies from $0.845$ to $1.25$. By combining this result with $c$'s dependence on the parameters $\sigma$ and $\zeta$, this leads to the conclusion that near the HB the wavelength of the spiral waves increases with $\sigma$ and decreases with $\zeta$, which was confirmed by our simulations (see, e.g., Fig.~\ref{diagram_phases}). 
 The prediction (\ref{lambda}) can be used to theoretically estimate the spiral wavelength, see e.g.~Fig.~\ref{EI_vs_BS} (left). As an example, the parameters used in Fig.~\ref{BSphase} correspond to $c\approx 0.8$ and $\epsilon\approx 0.255$, and therefore (\ref{lambda}) yields $\lambda_{{\rm CGLE}} \approx 27.1$ and a physical wavelength $\lambda \approx 27.1 / 0.255 \approx 106.3$. Yet, as the example in Fig.~\ref{BSphase} is not particularly close to the HB ($\epsilon\approx 0.255$), the wavelength found in the simulations is shorter than the prediction of  (\ref{lambda}). In the next section, we will see that a more accurate estimate accounting for the distance from the HB leads to $\lambda \approx 71.4$, which is in excellent agreement with the numerical  solutions of (\ref{pde}) as well as with the lattice simulations of the metapopulation model, see Fig.~\ref{BSphase}~(right). 

Fig.~\ref{lambda_vel} also shows that, near the HB, the spiral velocity varies little within the bound state phase, with values decaying from $v_{{\rm CGLE}}\approx 3.0$ to $v_{{\rm CGLE}}\approx 2.7$ when $c$ varies from $0.845$ to $1.25$ and $\delta=1$. It is worth noting that in a number of earlier works with $\mu=0$, the quantity  $v$ was considered to not vary with the CGLE parameter $c$, see, e.g., \cite{RMF,RF,rulands13}.

\begin{figure}
	\includegraphics[width=\linewidth]{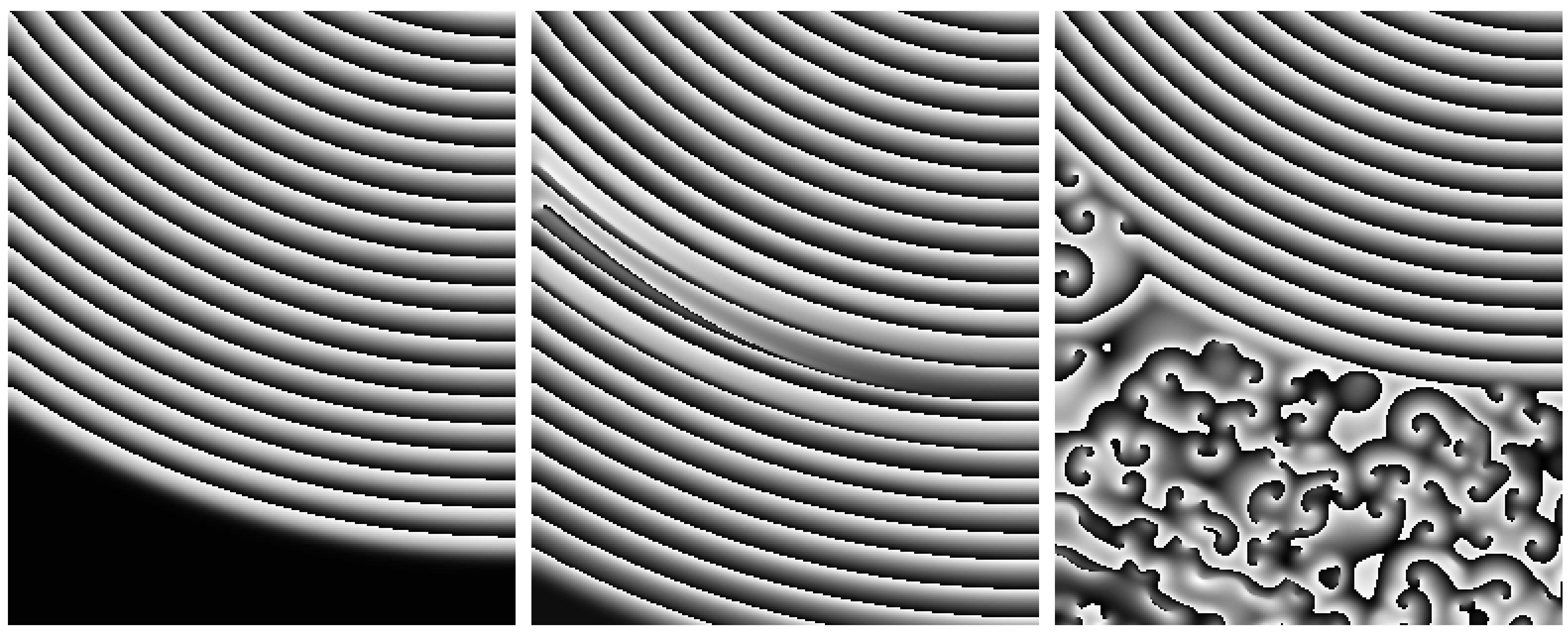}
	\caption{Space and time development of a spiral wave solution of the CGLE (\ref{CGLE}) with $c = 1.5$ and $\delta=1$ in the EI phase (argument of $\mathcal{A}$ encoded in grayscale): At time $t=700$ the spiral wave propagates with a wavelength $\lambda_{{\rm CGLE}}\approx 13.7$ (left). Subsequently, the arms start to deform ($t=800$, middle) and then a far--field break--up, due to a convective Eckhaus instability, occurs causing the spiral arms to break into an intertwining of smaller spirals ($t=900$, right), see text.}
	\label{EI-spiral}
\end{figure}

\begin{figure}
	\includegraphics[width=0.49\linewidth]{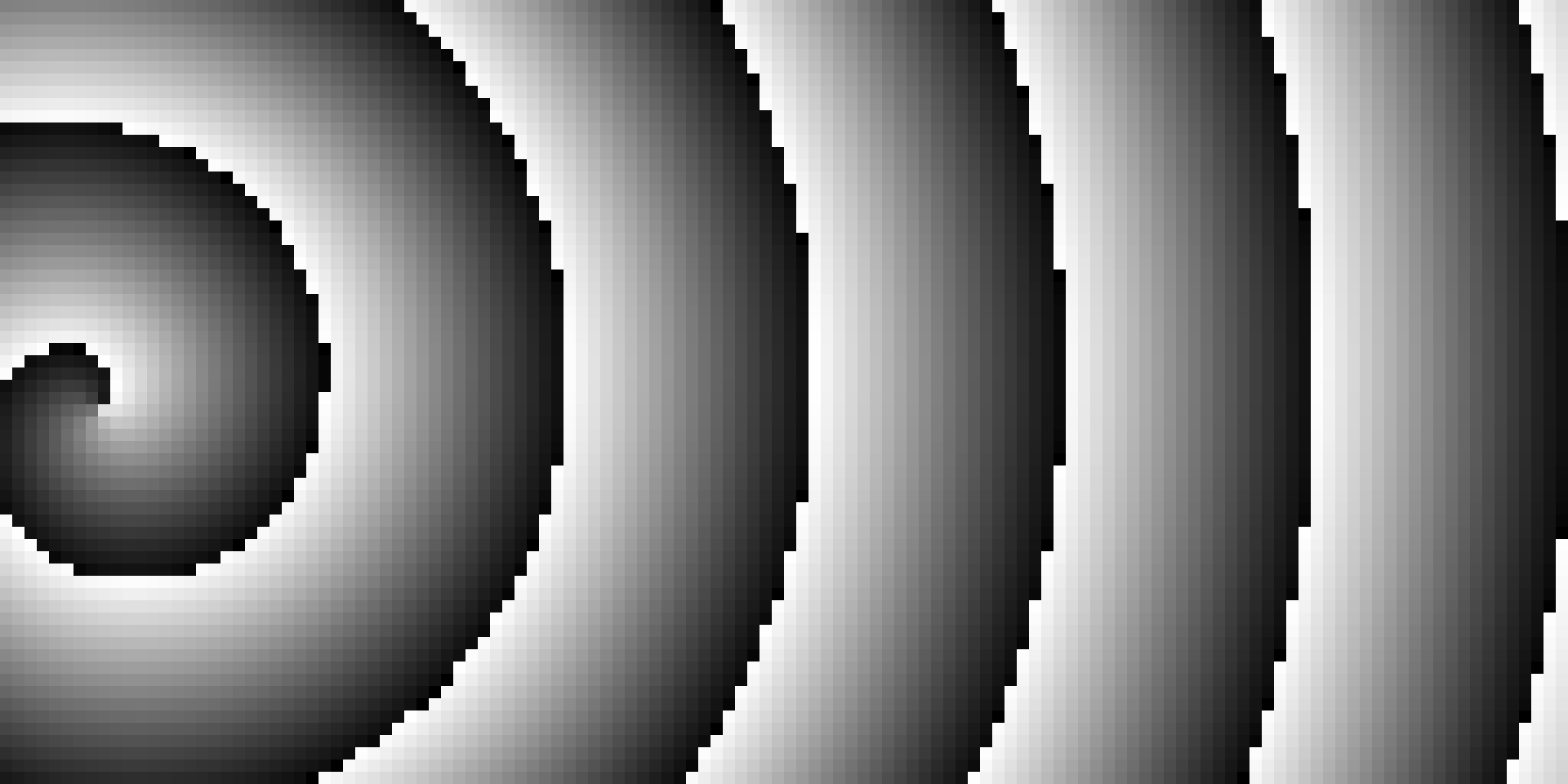}
	\includegraphics[width=0.49\linewidth]{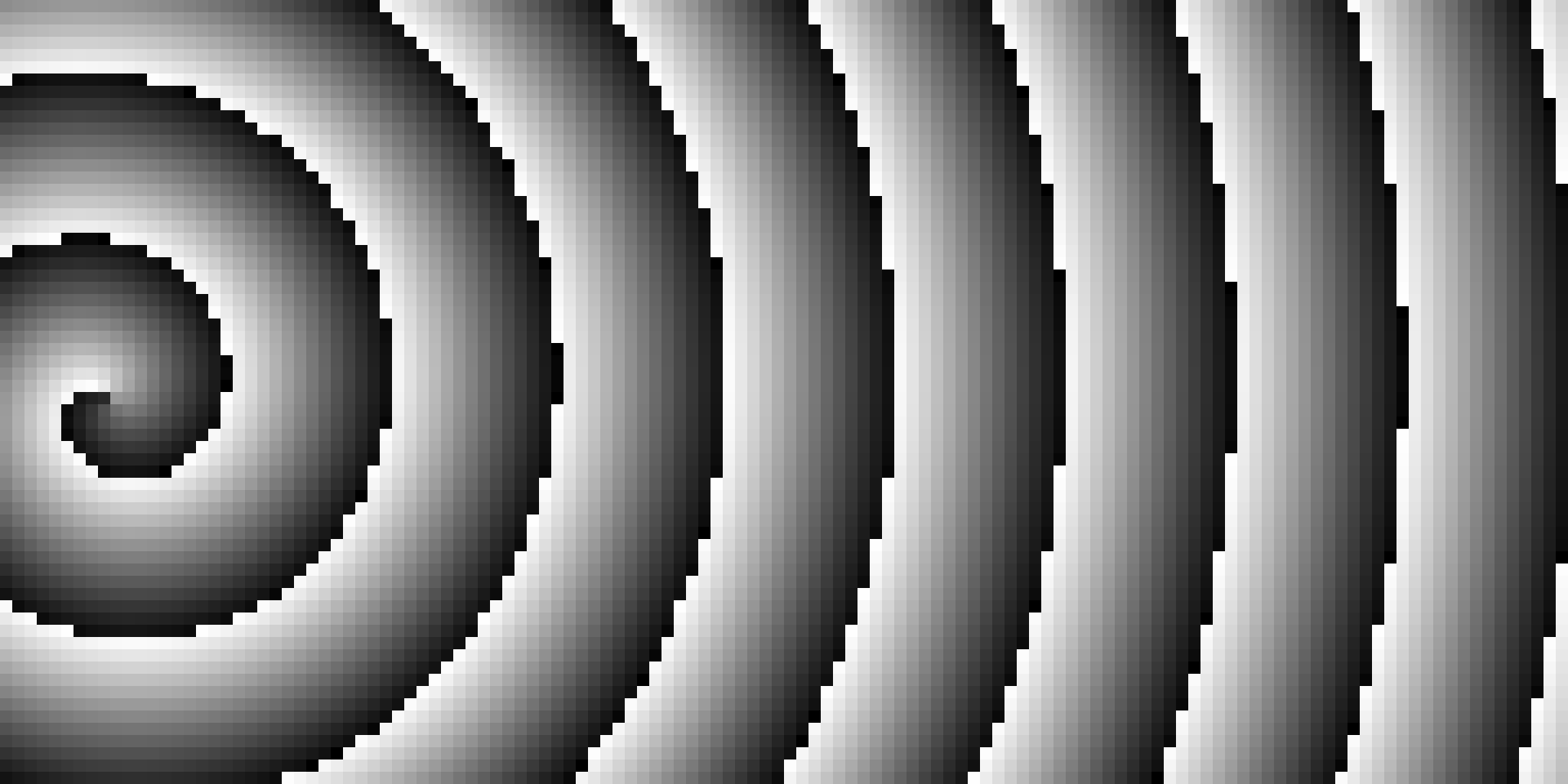}
	\caption{Wavelengths of well--developed spiral wave solutions of the CGLE (\ref{CGLE}) with $\delta=1$ in the BS and EI phases (argument of $\mathcal{A}$  encoded in grayscale). Here, the wavelengths are measured by counting pixels. Left: $c = 1.0$ and spirals are stable (BS phase). The measured wavelength is $20.2$ and compares well with the theoretical predictions $\lambda_{{\rm CGLE}}\approx 20.3$ obtained from (\ref{lambda}) with $|\mathcal{A}|^2\approx R^2$ measured as $0.904$. Right: $c = 1.5$ and spirals waves are in the EI phase, but their arms are still unperturbed. The Eckhaus instability will cause a far--field break--up further away from the core (not shown here, see text and Fig.~\ref{EI-spiral}). The measured wavelength of $13.8$ is in excellent agreement with $\lambda_{{\rm CGLE}}\approx 13.7$ from (\ref{lambda}) with $|\mathcal{A}|^2\approx R^2$ measured as $0.791$.}
	\label{EI_vs_BS}
\end{figure}

\subsection{Eckhaus instability phase ($1.25 \lesssim c \lesssim 1.75$)}
\label{SecEI}

As shown in Figs.~\ref{R2} and \ref{lambda_vel} the amplitude of the traveling wave solution (when it is valid) and the spirals' wavelength vary with $c$. As a consequence, the wavelength decreases when $c$ increases and above a critical value $c_{{\rm EI}}$ the spiral waves become unstable, see Fig.~\ref{EI-spiral}. Here, we   demonstrate the predictive power of our approach by deriving  $c_{{\rm EI}}$ from our controlled CGLE (\ref{CGLE}) and by characterizing the convective Eckhaus instability arising in the range $c_{{\rm EI}} < c < c_{{\rm AI}}$.

When $c_{{\rm EI}} < c < c_{{\rm AI}}$, small perturbations of the spiraling patterns, which normally decay for $c < c_{{\rm EI}}$ grow, and are convected away from the cores, this is the Eckhaus instability, as illustrated in Fig.~\ref{EI-spiral}. These instabilities eventually cause the far--field break--up of the spiraling patterns and the emergence of an intertwining of smaller spirals, see Fig.~\ref{EI-spiral} (rightmost). Before the far--field break--up occurs the properties of spirals far from the core are still well described by the plane wave solution of the CGLE (\ref{CGLE}) and the dispersion relation (\ref{disp-rel}). In particular, Fig.~\ref{EI_vs_BS} illustrates that the spiral wavelength relatively close to their cores (absence of far--field break--up), but still at a sufficient distance from them for the traveling wave ansatz to be valid, is in excellent agreement with the theoretical prediction (\ref{lambda}), see also Fig.~\ref{EI-spiral}~ (leftmost).

The convective nature of the instability makes it challenging to determine the critical value $c=c_{{\rm EI}}$  marking the onset of the Eckhaus instability, but its theoretical value can be predicted by considering a perturbation of the plane wave ansatz $\mathcal{A}=(1+\rho)R e^{i ({\bm k}.{\bm X}+\omega T+\varphi)}$ with $|\rho|,|\varphi|\ll 1$ as a solution of our CGLE (\ref{CGLE}). Substituting this expression into  (\ref{CGLE}) and seeking for a solution of the form $\rho \sim \varphi \sim e^{gT+i{\bm q}. {\bm X}}$~\cite{Hoyle}, we find that $\Re{(g)} > 0$ and the perturbation grows exponentially when $\delta k^2 > (3 + 2 c^2)^{-1}$, or equivalently when 
	\begin{eqnarray}
	\label{EIpert}
		R^2 < \frac{2 (1 + c^2)}{3 + 2 c^2}.
	\end{eqnarray}
In Fig.~\ref{R2}, the criterion (\ref{EIpert}) is used to determine the onset of the EI phase by plotting the measured $|\mathcal{A}|^2\approx R^2$ dependence on $c$ in the range $c=0.1 - 1.5$, yielding the estimate $c_{{\rm EI}} \approx 1.28$ that agrees well with the value $c_{{\rm EI}} \approx 1.25$ reported in the phase diagram of the two--dimensional CGLE \cite{CGLE}. The following condition on the spiral wavelengths in the physical domain of the PDEs (\ref{pde}) can be obtained from (\ref{lambda}) and (\ref{EIpert})
	\begin{eqnarray}
	\label{cond}
		\lambda < \frac{2\pi}{\epsilon} \sqrt{\delta(3 + 2 c^2)}.
	\end{eqnarray}
This gives an  upper bound $\lambda_{{\rm EI}} \approx 5 \pi \sqrt{\delta} / \epsilon$ for the spiral wavelength in the EI phase near the HB. We note that the wavelength in Fig.~\ref{EI-spiral} is indeed below $\epsilon \lambda_{{\rm EI}}$.

It is worth noting that for the model with $\mu=0, \delta_D=\delta_E$ and $\zeta=1$, the authors of Ref.~\cite{RF} observed the occurrence of an Eckhaus instability below a certain threshold $\sigma$ derived from an uncontrolled CGLE with $N=1$. We also note that our metapopulation model ($N \gg 1$) predicts not only the existence of Eckhaus instability but also an absolute instability phase at low values of $\sigma$, which has not been reported in Ref.~\cite{RF}.

\subsection{Spiral annihilation phase ($0< c \lesssim 0.845$)}

When $c < c_{{\rm BS}}$ near the HB, the spatio--temporal dynamics is characterized by the pair annihilation of colliding spirals. The phenomenon of spiral annihilation drives the system towards an homogeneous oscillating state filling the entire space in a relatively short time for low values of $c \ll c_{{\rm BS}}$. This phenomenon is not affected by fluctuations and not caused by any type of instabilities, but is a genuine nonlinear effect, and is predicted by the phase diagram of the two--dimensional CGLE~\cite{CGLE,SMR}. For this reason it has not been observed in studies of models, like those of Refs.~\cite{RMF,RF,rulands13}, not characterized by a Hopf bifurcation.

\begin{figure}
	\includegraphics[width=\linewidth]{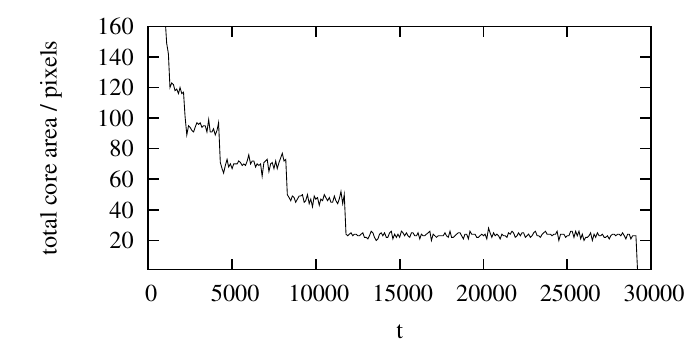}
	\caption{Staggered decay of the total core area in the solutions of the CGLE (\ref{CGLE}) with $c = 0.4$ and $\delta=1$. The initial condition consists of perturbations around $|\mathcal{A}|^2 = 0$. Here, after initial transients, 10 spirals remain with a total core area of approximately 120 pixels. Subsequently, further five annihilations occur marked by the sharp decreases in the total core area until the disappearance of all spirals.}
	\label{SA1}
\end{figure}

Theoretical results on the properties of the CGLE have established that in the spiral annihilation (SA) phase the stable equilibrium distance between two spirals increases asymptotically as the value of $c$ is lowered to $c_{{\rm BS}}$ which marks the end of the bound state phase \cite{CGLE}. In other words, unless the two spirals are separated by an infinite distance, they are destined to annihilate for any values $c < c_{{\rm BS}}$. The mean time necessary for the annihilation of two spirals separated by a certain distance, increases asymptotically as the value of $c$ approaches $c_{{\rm BS}}$ from below. At $c = c_{{\rm BS}}$ it takes an infinite time for the spirals to annihilate. 

\begin{figure}
	\includegraphics[width=\linewidth]{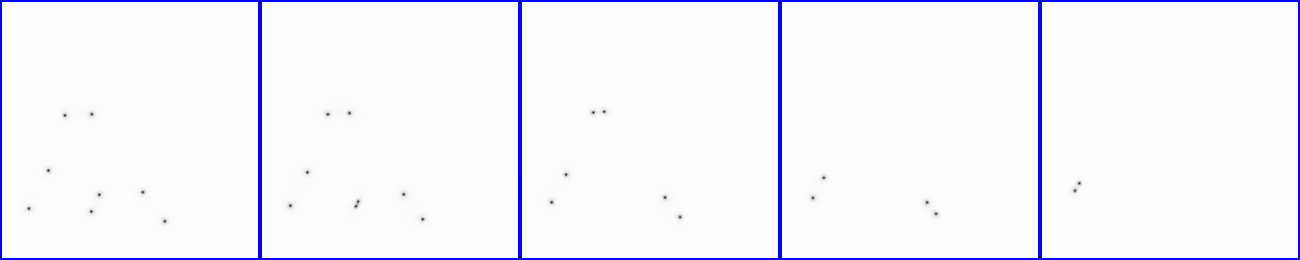}
	\caption{Spiral annihilation in the solutions of the CGLE (\ref{CGLE}) with $c = 0.1$ and $\delta=1$. The square modulus $|\mathcal{A}|^2$ is visualized here with dark pixels representing $|\mathcal{A}|^2 \approx 0$ while light pixels show regions where $|\mathcal{A}|^2 \approx 1$. Snapshots are taken at times $t = (1800, 2000, 2200, 2400, 2600)$ from left to right.}
	\label{SA2}
\end{figure}

An insightful way to characterize the SA phase consist of tracking the decay of the spiral core area in time. Spiral core area here refers to the number of points on the discrete grid forming the spiral core. To efficiently measure the spiral core area, we have used the modulus of the solution of the CGLE (\ref{CGLE}). We have confirmed that $|\mathcal{A}|^2$ is of order ${\cal O}(1)$ when there are traveling waves (see Figs.~\ref{R2} and \ref{CGLEarg}), but $|\mathcal{A}|^2$ drops rapidly to 0 within the small area of the core with such an area remaining approximately constant for a single core. The measure of the total core area is therefore a suitable quantity to characterize spiral annihilations. Practically, we have considered all points for which $|\mathcal{A}|^2 < 0.25$, as being part of spiral cores (dark pixels in Fig.~\ref{SA2}) and the total spiral core area is the number of all such points. We have also considered other limits such as $|\mathcal{A}|^2 < 0.1$ and $|\mathcal{A}|^2 < 0.5$ finding similar behavior for all cutoffs which are not too close to $1$. The actual value of the cutoff affects only the transients and not the long term dynamics dominated by the increasingly rare annihilation events. 

The spiral annihilations manifest themselves as sharp drops in the total core area equal to the area of the two colliding cores, as illustrated in Fig.~\ref{SA1} where the initial transient is characterized by a continuous decrease in the core area and the periods between first collisions are notably shorter since more spirals are present in the domain. Similarly, the time separating two successive annihilations takes always longer and the final annihilation takes longest (since spirals then need to cross the domain to collide and need to spin in opposite directions in order the annihilate). A visual representation of spiral annihilation for $c = 0.1$ is shown in Fig.~\ref{SA2} where $|\mathcal{A}|^2$ is coded in grayscale. Four pairs of dark spots, signifying the spiral cores with $|\mathcal{A}|^2 \approx 0$, are shown colliding and disappearing after approximately $3000$ time steps, which is an order of magnitude less than in Fig.~\ref{SA1} for $c = 0.4$. It has to be noted that the time to annihilation grows as $c$ approaches $c_{{\rm BS}}$ from below, as we confirmed in our simulations. While the spiral annihilation time tends to infinity when $c\to c_{{\rm BS}}$, here the closest value to $c_{{\rm BS}}$ that we considered was $c=0.4$  for which spiral annihilation typically occurs after a time exceeding $10^5$ time steps.

\begin{figure}
	\includegraphics[width=0.75\linewidth]{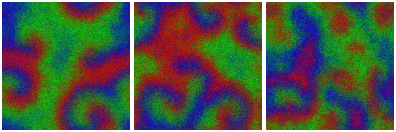}
	\caption{Spatial arrangements in the EI (left) and AI (center, right) phases as obtained from lattice simulations near the Hopf bifurcation. Parameters are $\sigma = \beta = \delta_E = \delta_D = 1$, $\mu = 0.02$, $L = 128$, $N = 64$, with $\zeta = 1.2$ in the EI phase (left) and $\zeta = (1.8, 2.4)$ in the AI phase (center, right). While the spatial arrangement is still characterized by (deformed) spiraling patterns in the EI phase, no spiraling arms can develop in the AI phase resulting in an incoherent spatial structure.}
	\label{AIEI}
\end{figure}

\subsection{Absolute instability phase ($ c \gtrsim 1.75$)}

When the value of the CGLE parameter exceeds $c > c_{{\rm AI}}\approx 1.75$ the instability occurring in the EI phase is no longer moving away from the core with the speed of the spreading perturbations exceeding the speed at which the spirals can convect them away. As illustrated in Fig.~\ref{AIEI}, when $c > c_{{\rm AI}}$, the perturbations grow locally destroying any coherent forms of spiraling patterns causing their absolute instability.

From the phase diagram Fig.~\ref{diagram_phases} we infer that the AI phase is the most extended phase (at least near the HB) and spiral waves are generally unstable when $\zeta \gg \sigma$, i.e. the rate of dominance-replacement greatly exceeds that of dominance-removal. This result can be compared with the absence of stable spiral waves reported in variants of the two-dimensional zero-sum model, see e.g. \cite{Matti} (where $N=1$ and $\sigma=\mu=0$).

\begin{figure}
	\includegraphics[width=\linewidth]{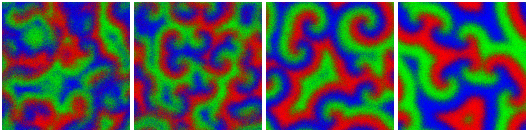}
	\caption{Four phases away from the HB (low mutation rate). Results of lattice simulations at low mutation rate $\mu=0.001 \ll \mu_H \approx 0.042$ (far away from the Hopf bifurcation) and with all the other parameters kept at same values as in Fig.~\ref{diagram_phases}. One recognises the AI, EI and BS phases (from left to right) while the spiral annihilation in the SA phase (rightmost panel) are no longer observed on the same length scales and time scales as in Fig.~\ref{diagram_phases}, see text.}
	\label{3Phases}
\end{figure}

\section{spatio--temporal patterns \& phases away from the Hopf bifurcation (low mutation rate)}
\label{awayHB}

While the spatio--temporal properties of the metapopulation model are accurately captured the CGLE (\ref{CGLE}) in the vicinity of the Hopf bifurcation (where $\epsilon$ is small), this is in principle no longer the case at low mutation rate $\mu$, when the dynamics occurs away from the Hopf bifurcation point. Yet, in this section we show how a qualitative, and even quantitative, description of the dynamics can be obtained from the CGLE (\ref{CGLE}) also when the mutation rate is low or vanishing, a case that has received significant attention in recent years~\cite{RMF,RF,Matti,He,lamouroux12,rulands13}.

\subsection{Phases and wavelengths at low mutation rate}

As  reported in Fig.~\ref{3Phases}, it appears that three of the four phases predicted by the CGLE (\ref{CGLE}) around the HB are still present far from the HB. Here, we first explore each of these phases. 
As illustrated in Figs.~\ref{3Phases} and \ref{AIEI}, when the rate $\zeta$ is decreased from a finite value to zero at fixed low mutation rate $\mu$ (with $\sigma,~\beta,~\delta_{D}$ and $\delta_{E}$ also kept fixed), the system is first in the absolute instability (AI),  then in the Eckhaus instability (EI) phase and eventually in the bound state (BS) phase. When $\zeta \gg \sigma$ and cyclic competition occurs mainly via dominance-replacement, AI in which spiral waves are unstable is the predominant phase, as observed in Refs.~\cite{Matti,He,SMR,FigshareMovies}. The EI and BS phases are also present near the HB and their common boundary is still qualitatively located as in the phase diagram of Fig.~\ref{diagram_phases}.  We have noted that, similarly to what happens near the HB, the onset of convective Eckhaus--like instability is accompanied by a decrease in the wavelength with respect to the BS phase and this  appears to hold even beyond the  regime of validity of  the CGLE approximation. The major effect on the phase diagram of lowering $\mu$ at fixed $\sigma$, when $\zeta$ is sufficiently low, is the replacement of the spiral annihilation phase by what appears to be an extended BS phase (see Fig.~\ref{3Phases}, rightmost): away from the HB and for low values of $\mu$ and $\zeta$, as in \cite{RMF}, instead of colliding and annihilating spiral waves turn out to be stable for the entire simulation time~\cite{SAnonHB}. However, it has also to be noted that when the dominance rate $\sigma$ considerably exceeds the other rates, an Eckhaus--like far--field break--up of the spiral waves occurs, see Sec.~\ref{awayHB}.B. 

\begin{figure}
	\includegraphics[width=\linewidth]{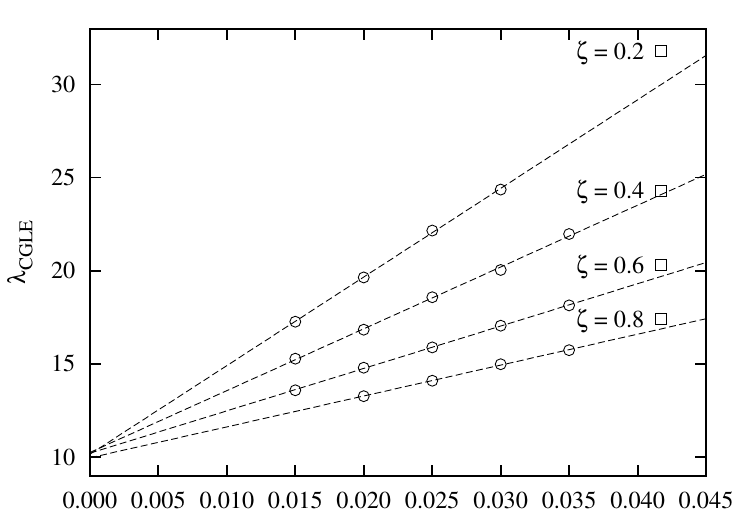}
	\centerline{$\mu$} % wavelength_convergence.eps has no label for x-axis
	\caption{Dependence of $\lambda_{{\rm CGLE}}=\epsilon \lambda $ on the vanishing mutation rate $\mu$ for various values of $\zeta$: Near the Hopf bifurcation $\mu\lesssim \mu_H\approx 0.042$, the wavelengths ($\square$) are obtained from the CGLE according to (\ref{lambda}). For lower values of $\mu$, the wavelengths ($\circ$) are measured in the solutions of (\ref{pde}), see text. When $\mu \to 0$, $\lambda $ approaches a value $ \widetilde{\lambda}(\sigma,\delta_D,\delta_E)$, see text. Parameters are: $\sigma=\beta=\delta_E=\delta_D=1$.}
	\label{Lambda_noHB}
\end{figure}

The AI, EI and BS phases at low mutation rates are characterized by the same qualitative properties as those studied in Sec.~\ref{nearHB}, (compare the upper panels of Fig.~\ref{diagram_phases} with Fig.~\ref{3Phases}). As a significant difference however, it has to be noted that the wavelength of the spiraling patterns in the BS and EI phases are shorter at low mutation rates than near the HB. To explore this finding we have studied how the wavelength depends on $\mu$. We have thus investigated how (\ref{lambda}) can be generalized at low values of $\mu$. To this end, the wavelengths of the spiral waves solutions of (\ref{pde}) were measured for $\mu$ ranging from $0.015$ to $0.035$ and for various values of $\zeta$ ($\sigma$ and $\beta=1$ are kept fixed). As shown in Fig.~\ref{Lambda_noHB}, the measured wavelength were compared with those obtained with (\ref{lambda}) when $ \mu= \mu_H$ and were found to be aligned and, quite remarkably, to collapse towards a single wavelength $\lambda \to \widetilde{\lambda}$ when $\mu=0$, where $\widetilde{\lambda}=\widetilde{\lambda}(\sigma,\delta_D,\delta_E)$ is a function of the non-mutation rates $\sigma,\delta_D,\delta_E$, with a slope that decreases when $\zeta$ is increased ($\beta$ is kept fixed). These results, summarized in Fig.~\ref{Lambda_noHB}, indicate that $\lambda$ depends linearly on $\mu$. Near $\mu\lesssim \mu_H$ the expression (\ref{lambda}) obtained from the CGLE (\ref{CGLE}) is a good approximation for the actual $\lambda$, whereas (\ref{lambda}) has to be rescaled by a linear factor, depending on $\sigma,\zeta$ and $\delta_{D,E}$, to obtain the wavelength when $\mu\approx 0$. 

The general effect of lowering $\mu$ is therefore to reduce $\lambda$ and hence to allow to fit more spirals in the finite system. As an example, the results reported in Fig.~\ref{Lambda_noHB} can be used together with  (\ref{lambda}) to accurately predict that the actual wavelength at $\mu = 0.02$ is $\lambda \approx  71.4$, which agrees excellently with what is found numerically (see Fig.~\ref{BSphase}).

\subsection{How does mobility and the rate of dominance influence the size of the spiraling patterns?}

Since we have introduced mobility by divorcing pair--exchange from hopping, yielding nonlinear diffusion in (\ref{pde}), we are interested in understanding how mobility influences the size of the spiraling patterns. 

\begin{figure}
	\includegraphics[width=\linewidth]{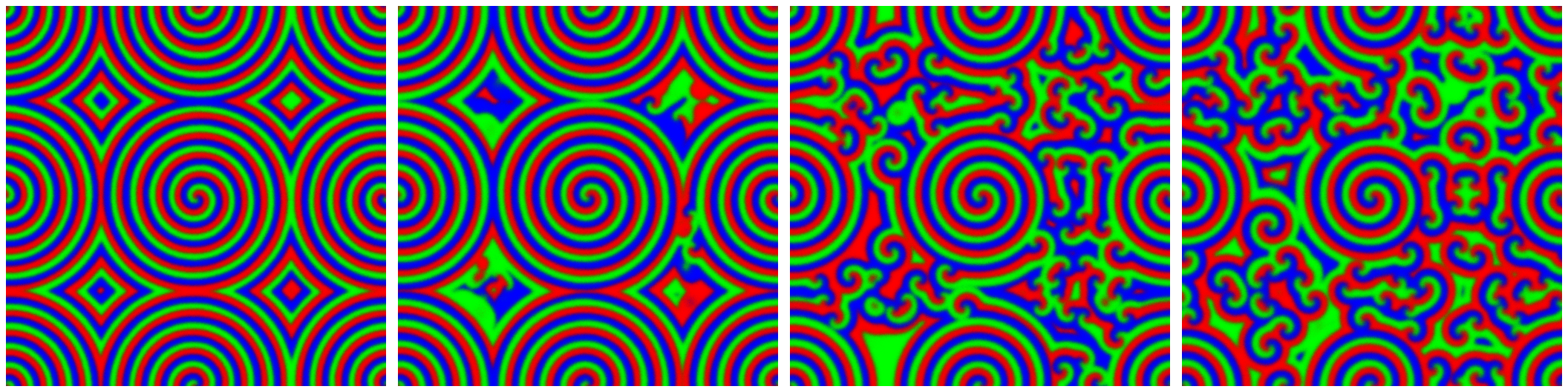}
	\caption{{\it (Color online)}. Effects of nonlinear mobility on spiraling patterns at zero mutation rate for various values of $\delta_D$ at $\delta_E$ fixed. Lattice simulations for the metapopulation model with $N=256,~L^2=512^2$, $\zeta=\mu=0$, $\sigma=\beta=1$, $\delta_E=0.5$, and $\delta_D=(0.5,1,1.5,2)$ from left to right. Spiral waves are stable and form geometric patterns when $\delta_D=\delta_E$ (leftmost, linear diffusion), Eckhaus--like instability occurs when $\delta_D>\delta_E$ and cause their far--field break--up resulting in a disordered intertwining of small spiraling patterns of short wavelengths, see text. }
	\label{nonlinmob}
\end{figure}

In Sec.~\ref{nearHB}, we have seen that only linear mobility, via an effective linear diffusion term in (\ref{CGLE}), matters near the HB. The latter does not influence the stability of the spiraling patterns but sets the spatial scale: changing the effective diffusion coefficient $\delta \to \alpha \delta$ ($\alpha>0$) rescales the space according to ${\bm x} \to {\bm x}/\sqrt{\alpha}$, as confirmed by numerical results. A more intriguing situation arises far from the HB, where the use of the CGLE is no longer fully legitimate: Nonlinear mobility is thus found to be able to alter the stability of the spiral waves (in addition to influence the spatial scale). As illustrated by Fig.~\ref{nonlinmob}, when the intensity of nonlinear mobility is increased (by raising $\delta_D$ at fixed $\delta_E$) in the BS phase, the spiral waves that were stable under linear diffusion (see Fig.~\ref{nonlinmob}, leftmost) disintegrate in an intertwining of spiral waves of limited size and short wavelength. It thus appears that nonlinear mobility promotes the far field breakup of spiral waves and enhances their convective instability via an Eckhaus--like mechanism resulting in a disordered intertwining of small spiraling patterns, see Fig.~\ref{nonlinmob} (rightmost). Furthermore, since the dominance--removal reaction is the only process that creates empty spaces that can be exploited by individuals for hopping onto neighboring patches, we expect that nonlinear mobility would be stronger at low value of $\sigma$ and for sufficiently high hopping rate $\delta_D$~\cite{inprep}.

\begin{figure}
	\includegraphics[width=\linewidth]{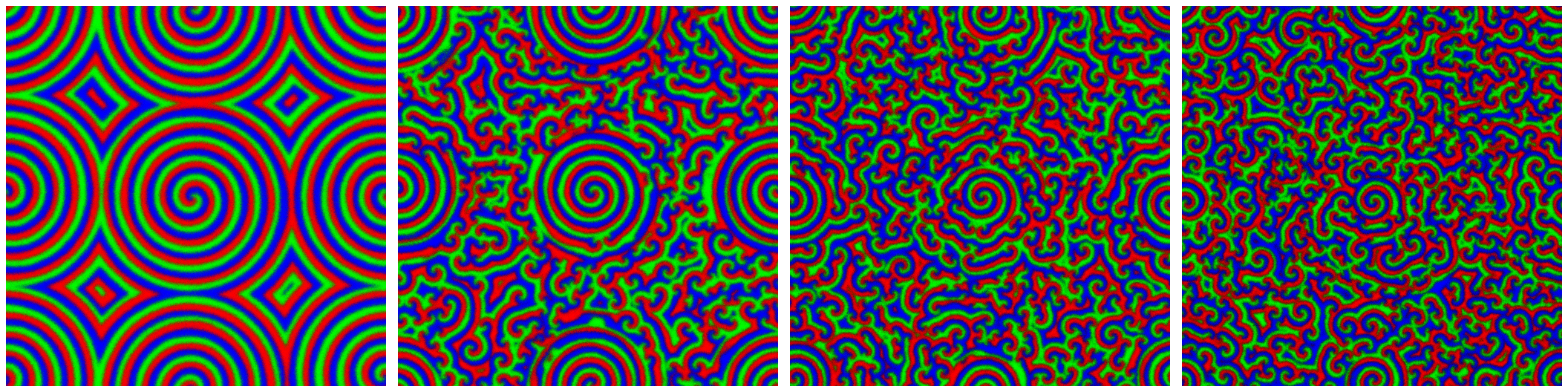}
	\caption{{\it (Color online)}. Raising $\sigma$ away from HB cause instability: Lattice simulations for the metapopulation model with $N=64, L^2=512^2$, $\zeta=\mu=0$, $\beta=1$, $\delta_D=\delta_E=0.5$, and $\sigma=(1,2,3,4)$ from left to right. While the spiral waves are stable and form a geometrically ordered when $\sigma=1$ (leftmost panel), Eckhaus--like instability occurs when $\sigma$ is raised and cause their far--field break--up (middle panels). When $\sigma=4$, the ordered spiraling patterns is entirely disintegrated and replaced by a disordered intertwining of spirals of small size and short wavelengths (rightmost panel), see text. }
	\label{raising_sigma}
\end{figure}

As already noticed in \cite{jiang11} for a version of the model (with $\zeta=\mu=0$, $\delta_D=\delta_E$ and $N=1$) considered here, it turns out that a similar mechanism destabilizes the spiral waves when the dominance--removal rate $\sigma$ is raised, with all the other parameters maintained fixed, as illustrated in Fig.~\ref{raising_sigma}. It indeed appears that spiral waves become far--field unstable after their wavelength have been reduced by raising $\sigma$. For high values of $\sigma$, any geometrically-ordered pattern is disintegrated into a disordered myriad of small intertwining spirals of reduced wavelength. It is noteworthy that the reduction of $\lambda$ as a result of raising $\sigma$ may seem counter-intuitive since the opposite happens near the HB (see Figs.~\ref{diagram_phases} and \ref{EI_vs_BS}), in accordance with the CGLE's predictions. As a possible explanation, we conjecture that the wavelength $\widetilde{\lambda}$ approached when $\mu$ vanishes is a decreasing function of $\sigma$.

So far, we have seen that the description in terms of (\ref{pde}) and their approximation by the CGLE (\ref{CGLE}) provide a faithful description of the spatio--temporal properties of the metapopulation model, which appear to be driven by nonlinearity rather than by noise when the carrying capacity is sufficient to allow a meaningful size expansion. However, when nonlinear mobility and/or the dominance--removal rates are high, the deterministic description in terms of (\ref{pde}) yield spiraling patterns of short wavelengths and limited size. In this case, the characteristic scale of the resulting patterns is too small to lead to coherent structures and, while the deterministic description (at high resolution) may predict a disordered intertwining of small spirals, demographic noise resulting from a low carrying capacity $N$ typically leads to noisy patches of activity on the lattice rather than to spiraling patterns~\cite{inprep}.

\section{Discussion \& Conclusion}

In this work, we have investigated the spatio--temporal patterns arising from the cyclic competition between three species in two dimensions. For this, we have considered a generic model that unifies the evolutionary processes considered in earlier works (e.g., in \cite{RMF,RF,Matti,He,rulands13}). Here, the rock--paper--scissors cyclic interactions between the species are implemented through dominance--removal and dominance--replacement processes. In addition to the cyclic competition, individuals can reproduce, mutate and move, either by swapping their position with a neighbor or by hopping onto a neighboring empty space, which yields nonlinear mobility. Inspired by recent experiments on microbial communities~\cite{Kerr,Nahum}, we have formulated a metapopulation model consisting of an array of patches of finite carrying capacity, each of which contains a well--mixed sub-population. While movement occurs between individuals of neighboring patches, all the other processes take place within each patch. The metapopulation formulation permits a neat description of the system's dynamics and provides an ideal setting to study the influence of nonlinearity and stochasticity. In particular, significant analytical progress is feasible in the vicinity of the Hopf bifurcation (HB) caused by the mutation process. 

By investigating the deterministic and stochastic descriptions of the system analytically and numerically, the main achievement of this work is to provide the detailed phase diagram of a generic class of spatial rock--paper--scissors games along with the comprehensive description of the spiraling patterns characterizing the various phases. Our main analytical approach relies on the model's complex Ginzburg--Landau equation (CGLE) derived from a multiscale perturbative expansion in the vicinity of the system's HB. As a major difference with respect to what was done in the vast majority of earlier works on this subject, our CGLE provides us with a fully {\it controlled} approximation of the dynamics around the bifurcation point. We have been able to exploit the well--known properties of the CGLE to obtain the accurate phase diagram near the HB in terms of a single parameter. The diagram is characterized by four phases, called ``absolute instability'' (AI), ``Echkaus instability'' (EI), ``spiral annihilation'' (SA) and ``bound state'' (BS). Spiral waves are found to be stable and convectively unstable in the BS and EI phases respectively, where their wavelength and velocity have been obtained from the dispersion relation of the CGLE and found to be in good agreement with results of both the deterministic and lattice simulations of the system. We have also been able to derive the threshold separating the BS and EI phases. The SA phase, whose existence is found to be limited to the vicinity of the HB, is characterized by the spiral waves' annihilation time (inferred from the CGLE). Finally, we have found that there is always a regime (AI phase), typically arising when dominance--replacement outcompetes dominance--removal, where any coherent form of spiraling patterns is prevented by growing local instabilities. We have also been able to take advantage of the CGLE to analyze the model's spatio--temporal properties at low mutation rates, i.e. far from the HB. In particular, we have found that at low mutation rate the AI, EI and BS phases are still present whereas the SA phase is replaced by what appears to be an extended BS phase. We have found that the wavelength of the spiral waves in the BS and EI phases decays linearly with the mutation rate. 
While we have focused on the two-dimensional system for its biological relevance, it worth noting that our analytical approach based on the CGLE is general and can also cover the cases of one and three spatial dimensions: One would then obtain different phase diagrams where in which one would notably find traveling waves (in one dimension) and scroll waves (in three dimensions) instead of instead of spiraling patterns.

In general, we have seen that phenomena like far--field break--up and convective instabilities that characterize the EI phase, and limit the size of the spirals as well as their coherent arrangement can also be caused by nonlinear mobility and by high dominance--removal rate. Under high nonlinear mobility or for high dominance--removal rate, the system may exhibit spiraling patterns of short wavelength and limited size even in the extended BS phase. In this case, if the carrying capacity is low, the intensity of demographic noise may prevent the visualization of spiraling patterns on the discrete lattice~\cite{inprep}. 

Our findings shed further light on the spatio--temporal arrangement of population in cyclic competition and provide possible explanations for the lack of observation of spiraling patterns in microbial experiments as those of Ref.~\cite{Kerr}. One possible explanation could be that the experimental parameters would correspond to a regime where spiral waves are unstable. Another plausible explanation could be that the time scale on which the experiments of Ref.~\cite{Kerr} have been carried out (several days) is much shorter than the time necessary for the formation of spiraling patterns in the simulations of the model. This would imply that spiraling patterns would take very long (perhaps several months) to form on a Petri dish, which might explain why they have remained elusive. 

\section{acknowledgments}
BS is grateful for the support of an EPSRC PhD studentship (Grant No. EP/P505593/1).

\appendix

\section{Stochastic dynamics \& van Kampen size expansion}
\label{AppendixA}

In this appendix, we explain how the stochastic dynamics of the generic metapopulation model (\ref{sel})-(\ref{migr})
can be captured by the system's master equation. We also outline how the latter can be expanded to yield 
a more amenable description of the dynamics~\cite{SizeExp}.

\subsection{Master Equation}

We here derive the master equation (ME) governing the stochastic dynamics of the generic metapopulation model. Combining the reaction rates with appropriate combinatorial factors, the transition probabilities for each intra--patch reactions 
(\ref{sel}) and (\ref{non-sel}) can be written as
	\begin{align}
		T_i^\beta(\ensuremath{\boldsymbol\ell}) &= \beta \frac{N_{S_i}(\ensuremath{\boldsymbol\ell}) N_{\O}(\ensuremath{\boldsymbol\ell})}{N^2} \\
		T_i^\sigma(\ensuremath{\boldsymbol\ell}) &= \sigma \frac{N_{S_i}(\ensuremath{\boldsymbol\ell}) N_{S_{i+1}}(\ensuremath{\boldsymbol\ell})}{N^2} \\
		T_i^\zeta(\ensuremath{\boldsymbol\ell}) &= \zeta \frac{N_{S_i}(\ensuremath{\boldsymbol\ell}) N_{S_{i+1}}(\ensuremath{\boldsymbol\ell})}{N^2} \\
		T_i^\mu(\ensuremath{\boldsymbol\ell}) &= \mu \frac{N_{S_i}(\ensuremath{\boldsymbol\ell})}{N}.
	\end{align}
The combinatorial factors, such as $N_{S_{i}}(\ensuremath{\boldsymbol\ell}) N_{S_{i+1}}(\ensuremath{\boldsymbol\ell}) / N^2$, express the probability of species $S_{i}$ and $S_{i+1}$ to interact within a patch at site $\ensuremath{\boldsymbol\ell}$. The same applies to $N_{S_i}(\ensuremath{\boldsymbol\ell}) N_{\O}(\ensuremath{\boldsymbol\ell}) / N^2$ for the probability of species $S_{i}$ encountering an empty space denoted by $\O$. Migration between two
neighboring patches occurs by pair--exchange (with rate $\delta_E$) and by hopping (with rate $\delta_D$)
 according to (\ref{migr}), which similarly yields the transition probabilities
	\begin{align}
		D_{i}^{\delta_D}(\ensuremath{\boldsymbol\ell}, \ensuremath{\boldsymbol\ell}') &= \delta_D \frac{N_{S_i}(\ensuremath{\boldsymbol\ell}) N_{\O}(\ensuremath{\boldsymbol\ell}')}{N^2} \\
		D_{i}^{\delta_E}(\ensuremath{\boldsymbol\ell}, \ensuremath{\boldsymbol\ell}') &= \delta_E 
\frac{N_{S_i}({\bm \ell})N_{S_{i\pm 1}}({\bm \ell}')}{N^2}.
	\end{align}
At this point, it is useful to introduce the one step-up and one step-down operators~\cite{SizeExp}. These act on a given state or transition by changing the numbers of individuals by $\pm1$, i.e. $\mathbb{E}_{i}^\pm N_{S_i}(\ensuremath{\boldsymbol\ell})=N_{S_i}(\ensuremath{\boldsymbol\ell})\pm 1$ and therefore
	\begin{equation}
		\mathbb{E}_{i}^\pm(\ensuremath{\boldsymbol\ell}) ~ T_{i}^\beta(\ensuremath{\boldsymbol\ell})
			= \beta \frac{(N_{S_i}(\ensuremath{\boldsymbol\ell}) \pm 1) N_{\O}(\ensuremath{\boldsymbol\ell})}{N^2}.
	\end{equation}
 This allows the total transition operator for intra--patch reactions to be written as
	\begin{align}
		\mathbb{T}_{i}(\ensuremath{\boldsymbol\ell})
		&= \left[ \mathbb{E}_{i+1}^+(\ensuremath{\boldsymbol\ell}) - 1 \right] T_{i}^{\sigma}(\ensuremath{\boldsymbol\ell}) \nonumber \\
		&+ \left[ \mathbb{E}_{i}^-(\ensuremath{\boldsymbol\ell}) \mathbb{E}_{i+1}^+(\ensuremath{\boldsymbol\ell}) - 1 \right] T_{i}^{\zeta}(\ensuremath{\boldsymbol\ell}) \\
		&+ \left[ \mathbb{E}_{i}^-(\ensuremath{\boldsymbol\ell}) - 1 \right] T_{i}^{\beta}(\ensuremath{\boldsymbol\ell}) \nonumber \\
		&+ \left[ \mathbb{E}_{i}^-(\ensuremath{\boldsymbol\ell}) \mathbb{E}_{i+1}^+(\ensuremath{\boldsymbol\ell}) + \mathbb{E}_{i}^-(\ensuremath{\boldsymbol\ell}) \mathbb{E}_{i-1}^+(\ensuremath{\boldsymbol\ell}) - 2 \right] T_{i}^{\mu}(\ensuremath{\boldsymbol\ell}). \nonumber
	\end{align}
The general form of the terms $[\mathbb{E}_{\dots}^{\pm} - 1] T_{\dots}^{\dots}$ originates from the gain and loss terms in probability to find the system in a particular state. Correspondingly, the total migration operator for diffusions between neighboring subpopulations reads
	\begin{align}
\label{equ:DDi}
		&\mathbb{D}_i(\ensuremath{\boldsymbol\ell},\ensuremath{\boldsymbol\ell}') 
		= \left[
			\mathbb{E}_{i}^+(\ensuremath{\boldsymbol\ell})
			\mathbb{E}_{i}^-(\ensuremath{\boldsymbol\ell}')
		- 1 \right] D_{i}^{\delta_D}(\ensuremath{\boldsymbol\ell}, \ensuremath{\boldsymbol\ell}') \\
		&\quad + \left[
			\mathbb{E}_{i}^+(\ensuremath{\boldsymbol\ell}) \mathbb{E}_{i\pm 1}^-(\ensuremath{\boldsymbol\ell})
			\mathbb{E}_{i}^-(\ensuremath{\boldsymbol\ell}') \mathbb{E}_{i\pm 1}^+(\ensuremath{\boldsymbol\ell}')
		- 1 \right] D_{i}^{\delta_E}(\ensuremath{\boldsymbol\ell}, \ensuremath{\boldsymbol\ell}'). \nonumber
	\end{align}
Finally, we can write the master equation for the probability $P(\bm{N},t)$ of a system occupying a certain state $\bm{N}$ at time $t$ by summing the operators over all species $i \in \{1, 2, 3\}$ and subpopulations $\ensuremath{\boldsymbol\ell} \in \{1, \dots, L\}^2$, which yields
	\begin{equation}
	\label{equ:master}
		\frac{dP(\bm{N},t)}{dt} =  \sum_{i=1}^3 \sum_{\ensuremath{\boldsymbol\ell}}^{L \times L}
		\left[ \mathbb{T}_{i}(\ensuremath{\boldsymbol\ell}) + \frac{1}{2}\sum_{\pm}\sum_{\ensuremath{\boldsymbol\ell}' \in \ensuremath{\boldsymbol\ell}} \mathbb{D}_{i}(\ensuremath{\boldsymbol\ell}, \ensuremath{\boldsymbol\ell}') \right] P(\bm{N},t).
	\end{equation}
Here, the term $\ensuremath{\boldsymbol\ell}' \in \ensuremath{\boldsymbol\ell}$ indicates summation over all neighbors of patch $\ensuremath{\boldsymbol\ell}$ and $\sum_{\pm}$ denotes the sum over $i\pm 1$ in (\ref{equ:DDi}). In addition, $\bm{N}=\{N_{\O}(\ensuremath{\boldsymbol\ell}), N_{S_i}(\ensuremath{\boldsymbol\ell})| i=1,2,3, \ensuremath{\boldsymbol\ell}\in L\times L\}$ is defined as a collection of all $N_{S_i}(\ensuremath{\boldsymbol\ell})$'s and empty spaces $N_{\O}(\ensuremath{\boldsymbol\ell})$ in all subpopulations specifying uniquely the state of the entire system. Later, $\bm{\eta}$ is used to symbolise a similar collection for fluctuations $\eta_i(\ensuremath{\boldsymbol\ell})$ defined below.

\subsection{System Size Expansion}
\label{sec:system_size_expansion}

While  the mathematical treatment of (\ref{equ:master}) represents a formidable problem, significant progress can be made by performing an expansion 
in the inverse of the carrying capacity $N$~\cite{SizeExp}. Such a system size expansion requires the introduction of new rescaled variables. The normalised abundances (densities) of species are equal to $s_{i}(\ensuremath{\boldsymbol\ell}) = N_{i}(\ensuremath{\boldsymbol\ell}) / N$. Here, for convenience the dependence on $\ensuremath{\boldsymbol\ell}$ is dropped, and the fluctuations $\eta_{i}(\ensuremath{\boldsymbol\ell})$ around the fixed point $s^*$ are defined to scale with $\sqrt{N}$ such that
	\begin{equation}
		\eta_{i}(\ensuremath{\boldsymbol\ell}) = \sqrt{N} \left(s^* - s_{i}(\ensuremath{\boldsymbol\ell}) \right), \quad \mbox{where} \; s^*=\frac{\beta}{\beta+3\sigma},
	\end{equation}
which after differentiating with respect to time becomes
	\begin{equation}
		\label{equ:van_kampen_ansatz}
		\frac{d \eta_{i}(\ensuremath{\boldsymbol\ell})}{dt} = -\sqrt{N} \frac{d s_{i}(\ensuremath{\boldsymbol\ell})}{dt}.
	\end{equation}
With this assumption, it is now possible to write the master equation for a (redefined) probability density $\Pi(\bm{\eta}, t)$ in terms of the fluctuations $\eta_{i}(\ensuremath{\boldsymbol\ell})$.
As usual, the
time is rescaled as $t \rightarrow t / N$ and the left hand side of \eqref{equ:master} thus  becomes
	\begin{eqnarray}
		\frac{1}{N}\frac{\partial\Pi(\bm{\eta}, t)}{\partial t} 
			 - \sum_{i=1}^3 \sum_{\ensuremath{\boldsymbol\ell}}^{\{1,\dots,L\}^2}
				\frac{1}{\sqrt{N}} \frac{d s_{i}(\ensuremath{\boldsymbol\ell})}{dt} \frac{\partial\Pi(\bm{\eta}, t)}{\partial\eta_{i}(\ensuremath{\boldsymbol\ell})}.
	\end{eqnarray}
The right hand side of \eqref{equ:master} can be written in a similar way by introducing $s_{i}(\ensuremath{\boldsymbol\ell})$ and $\eta_{i}(\ensuremath{\boldsymbol\ell})$ variables. The step up and step down operators are also expanded in their differential form which, up to the order $\mathcal{O}(N^{-1})$, reads
	\begin{equation}
		\mathbb{E}_{i}^{\pm}(\ensuremath{\boldsymbol\ell}) = 1 \pm
			\frac{1}{\sqrt{N}}
				\frac{\partial}{\partial \eta_{i}(\ensuremath{\boldsymbol\ell})} +
			\frac{1}{2} \frac{1}{N}
				\frac{\partial^2} {\partial \eta_{i}^2(\ensuremath{\boldsymbol\ell})}.
	\end{equation}
The results of successive application of the operators can be obtained by multiplying their differential forms. For example, the application of $\mathbb{E}_{i}^+(\ensuremath{\boldsymbol\ell}) \mathbb{E}_{i}^-(\ensuremath{\boldsymbol\ell}')$ results in
	\begin{align}
		\mathbb{E}_{i}^+(\ensuremath{\boldsymbol\ell}) \mathbb{E}_{j}^-(\ensuremath{\boldsymbol\ell}') &= 1 +
			\frac{1}{\sqrt{N}}
				\left(
					\frac{\partial}{\partial \eta_{i}(\ensuremath{\boldsymbol\ell})} - 
					\frac{\partial}{\partial \eta_{j}(\ensuremath{\boldsymbol\ell}')}
				\right) \nonumber \\
			&+ \frac{1}{2} \frac{1}{N}
				\left(
					\frac{\partial}{\partial \eta_{i}(\ensuremath{\boldsymbol\ell})} -
					\frac{\partial}{\partial \eta_{j}(\ensuremath{\boldsymbol\ell}')}
				\right)^2.
	\end{align}
After some algebra, the terms at the same order of $N$ can be collected on both sides of the master equation \eqref{equ:master}. At order $\mathcal{O}\big( N^{-1/2} \big)$, the leading terms describe the time evolution of the species densities $s_{i}(\ensuremath{\boldsymbol\ell})$. Leaving out the migration terms for now and collecting all intra--patch reaction terms, the ordinary differential equations describing changes in one patch can be written down. These mean field equations are also referred to as the rate equations. Since only the subpopulation in one patch is considered at this point and space is currently irrelevant, the spatial variable $\ensuremath{\boldsymbol\ell}$ in $s_{i} (\ensuremath{\boldsymbol\ell})$ is temporarily dropped. With the introduction of $\bm{s} = (s_1, s_2, s_3)$ and $r = s_1 + s_2 + s_3$, the ODEs read
	\begin{align}
		\label{equ:replicator}
		\frac{d s_i}{d t} &= s_i [\beta (1-r) - \sigma s_{i-1} + \zeta (s_{i+1} - s_{i-1})] \nonumber \\
		&+ \mu (s_{i-1} + s_{i+1} - 2 s_i) = \mathcal{F}_i(\bm{s}),
	\end{align}
which corresponds to the mean field rate equations (\ref{RE}).

When migration terms are accounted for, the size expansion to order $\mathcal{O}\big(N^{-1/2} \big)$ yields terms that
describe the deterministic spatial dynamics of the model. In the suitable continuum limit, these lead to the following partial differential equations (PDEs) for the continuous coordinate $\bm{x}={\cal S}({\bm \ell}/L)$ describing the system's dynamics on a domain of size ${\cal S}$:
	\begin{align}
		\label{equ:pdes}
		\frac{\partial {s}_{i}(\bm{x})}{\partial t} &= \mathcal{F}_i(\bm{s}(\bm{x})) + \delta_D~\left(\frac{{\cal S}}{L}\right)^2 \Delta s_{i}(\bm{x}) \\
			&+ ( \delta_D - \delta_E)~\left(\frac{{\cal S}}{L}\right)^2 \Big( s_{i}(\bm{x}) \Delta r(\bm{x}) - r (\bm{x}) \Delta s_{i}(\bm{x}) \Big),  \nonumber
	\end{align}
where $\mathcal{F}_i(\bm{s}(\bm{x}))$ in the first line coincides with the right-hand side of (\ref{equ:replicator}) where the spatial dependence of the densities is reinstated according to $s_i \to  s_{i}(\bm{x})$. At this point, it is useful to comment on the derivation and interpretation of (\ref{equ:pdes}), which coincides with (\ref{pde}). To lowest order,  the size expansion of the master equation with migration yields terms like $\delta_D \left[\sum_{\bm{\ell}' \in \bm{\ell}}s_{i}(\bm{\ell}')-4s_{i}(\bm{\ell})\right]$, where  $\bm{\ell}'$ are the four nearest--neighbor to site $\bm{\ell}$. To obtain the deterministic description of the model in the continuum limit on a domain of fixed size ${\cal S}\times {\cal S}$, we consider the number of lattice sites $L\to \infty$. In terms of the variable ${\bm x}=(x_1,x_2)$, the mobility rates of (\ref{migr}) are thus rescaled according to $\delta_{D,E} \to  \delta_{D,E} \left(\frac{{\cal S}}{L}\right)^2$ and  interpreted as diffusion coefficients. Therefore, in the continuum limit $\delta_D \left[\sum_{\bm{\ell}' \; {\rm n.n.} \; \bm{\ell}}s_{i}(\bm{\ell}')-4s_{i}(\bm{\ell})\right]\to \delta_D (\frac{{\cal S}}{L})^2 \Delta s_{i}(\bm{x})$, where the differential operator $\Delta=\partial_{x_1}^2 + \partial_{x_2}^2$ is the usual two--dimensional Laplacian. For the sake of comparison with lattice simulations, we set the domain size to be equal to the lattice size, i.e. ${\cal S}=L$ so the diffusion coefficients coincide with the mobility rates.
It is important to note that apart from the nonspatial ODE $ \mathcal{F}_i(\bm{s}(\bm{x}))$ \eqref{equ:replicator} and a linear diffusive term $\delta_D \Delta s_{i}(\bm{x})$ there are also additional nonlinear diffusive terms appearing in the second line of (\ref{equ:pdes}). These vanish only in the case of $\delta_D = \delta_E$ considered in the vast majority of other studies, e.g. in Refs~\cite{RMF,RF,Matti,rulands13}.

\section{Multiscale expansion and complex Ginzburg--Landau equation}
\label{AppendixB}
%
%\section{Appendix B: Multiscale expansion and complex Ginzburg--Landau equation}

In this appendix, we provide details of the  multi-scale asymptotic expansion leading to the complex Ginzburg--Landau equation (\ref{CGLE}) which provides a controlled (perturbative) approximation of the model's dynamics
in the vicinity of the Hopf bifurcation.

\subsection{Linear Transformations}
\label{sec:linear_transformations}

Before performing the asymptotic expansion can be performed, it is convenient to work with the shifted variables ${\bm u}=(u_1({\bm x}),u_2({\bm x}),u_3({\bm x}))={\bm M}({\bm s}-{\bm s}^*)$, where 
\begin{eqnarray}
	\label{M}
		{\bm M} =\frac{1}{\sqrt{6}}
		\left(
		\begin{array}{ccc}
			-1 & -1 & -2 \\
			-\sqrt{3} & \sqrt{3} & 0 \\
			\sqrt{2} & \sqrt{2} & \sqrt{2}
		\end{array}
		\right).
	\end{eqnarray}
With this transformation, the  origin coincides with
  the fixed point $\bm{s}^*$. In these new variables, the
 linear part of the rate equations \eqref{equ:replicator}  are in the Jordan normal form:
	\begin{equation}
		\frac{d \bm{u}(\bm{x}) }{d t} =
		\left[
		\begin{array}{ccc}
			\epsilon & -\omega_H & 0 \\
			\omega_H &  \epsilon & 0 \\
			0 & 0 & -\beta
		\end{array}
		\right]
		\bm{u}(\bm{x}),
	\end{equation}
where $\beta$ is the reproduction rate, $\omega_H = \frac{\sqrt{3}\beta (\sigma + 2\zeta)}{2(3\beta + \sigma)}$, $\epsilon=\sqrt{3(\mu_H -\mu)}$, and $\mu_H=\frac{\beta\sigma}{6(3\beta +\sigma)}$.
One notices that $u_3(\bm{x})$ decouples from  the oscillations in the $u_1(\bm{x})$-$u_2(\bm{x})$  at Hopf frequency $\omega_H$. The dynamics of three species abundances is therefore confined to two dimensions, which simplifies the multiscale expansion.

\subsection{Asymptotic Expansion}
\label{sec:asymptotic_expansion}

Once the linear transformation (\ref{M}) is performed onto
\eqref{equ:pdes}, we are interested in small perturbations of magnitude $\epsilon$ around the Hopf bifurcation by writing~\cite{Miller}
	\begin{equation}
		\label{equ:mu_epsilon}
		\mu = \mu_H - \frac{1}{3}\epsilon^2.
	\end{equation}
Unlike the strained coordinate method, the expansion assumes a general undetermined functional dependence on the new multiscale coordinates. As well-established in the theory of weakly nonlinear systems~\cite{Manneville,BenderOrszag}, the first step of the derivation is the multiscale expansion of time and space coordinates, e.g. $\partial_t \rightarrow \partial_t  + \epsilon^2 \partial_T$ and $\partial_x \rightarrow \epsilon \partial_X$ in one spatial dimension. The new coordinates $T = \epsilon^2 t$ and $\bm{X} = \epsilon \bm{x}$ are called  ``slow'' coordinates.  Therefore, the Laplace operator of \eqref{equ:pdes} becomes $\Delta \rightarrow \epsilon^2 \Delta_{\bm{X}}$ and is defined as $\Delta_{\bm{X}} = \partial^2_{X_1} + \partial^2_{X_2}$.
Furthermore, the variable $\bm{u}(\bm{x},t)$ is expanded in the perturbation parameter $\epsilon$. The expansion, up to the order $\mathcal{O}(\epsilon^3)$ where the CGLE is expected to appear, reads
	\begin{equation}
		\label{perturbation}
		 \bm{u}(\bm{x},t) = \sum_{n=1}^{3} \epsilon^n \bm{U}^{(n)}(t,T,\bm{X}).
	\end{equation}
As a results of these expansions, all scaling in $\epsilon$ is made explicit with the variables $T$, $\bm{X}$ and $\bm{U}^{(n)}$ for all $n$, being of order $\mathcal{O}(1)$.

Using the chain rule with the multiscale variables two times with $t$, $T=\epsilon^2 t$ and similarly for $\Delta u_i(\bm{x},t)$  with $\bm{X}=\epsilon \bm{x}$ results in a hierarchy of simple equations which can be solved at different orders of $\epsilon$ with necessary removals of the secular terms. These unbound terms arise naturally when the perturbation theory is applied to weakly nonlinear problems and their removal gives additional information about the system dynamics. Moreover, the Jordan normal form suggests that the first two components of $\bm{U}^{(n)}(t,T,\bm{X})$ should be combined into a complex number
	\begin{equation*}
		\mathcal{Z}^{(n)}(t,T,\bm{X}) = U_1^{(n)}(t,T,\bm{X}) + iU_2^{(n)}(t,T,\bm{X}).
	\end{equation*}
The hierarchy of simplified equations begins at the leading order $\mathcal{O}(\epsilon)$ where the first set of the equations reads
	\begin{align*}
		\partial_t \mathcal{Z}^{(1)}(t,T,\bm{X}) &= i \omega_H  \mathcal{Z}^{(1)}(t,T,\bm{X}) \\
		\partial_t U_3^{(1)}(t,T,\bm{X}) &= - \beta U_3^{(1)}(t,T,\bm{X})
	\end{align*}
These equations suggest oscillating and decaying solutions with the following ansatz proposed
	\begin{align*}
		\mathcal{Z}^{(1)}(t,T,\bm{X}) &= \mathcal{A}^{(1)}(T,\bm{X})e^{i\omega_H t} \\
		U_3^{(1)}(t,T,\bm{X}) &= 0.
	\end{align*}
where $\mathcal{A}^{(1)}(T,\bm{X})$ is the complex amplitude modulation at the ``slow'' time and length scales. Here, $U_3^{(1)}(t,T,\bm{X}) = 0$ is assumed as evident from the exponential decay with rate $\beta > 0$. 
At order 
$\mathcal{O}(\epsilon^2)$ one obtains $U_3^{(2)}=\frac{\sigma}{2\sqrt{3}\beta}|\mathcal{Z}^{(1)}|^2$, which corresponds to
the leading term for the invariant manifold considered in  \cite{RMF}. Continuing this procedure to  order $\mathcal{O}(\epsilon^3)$, a secular term is encountered. Canceling such a term yields the CGLE for $\mathcal{A}^{(1)}(T,\bm{X})$~\cite{CGLE}, which can be written as
	\begin{equation}
		\label{equ:cgle_two_c}
		\partial_{T}\mathcal{A}^{(1)} = \delta \Delta_{\bm{X}} \mathcal{A}^{(1)} + \mathcal{A}^{(1)} - (c_{r} + i c_{i}) |\mathcal{A}^{(1)}|^2 \mathcal{A}^{(1)}
	\end{equation}
where the constants in the coefficient of the ``cubic'' $|\mathcal{A}^{(1)}|^2 \mathcal{A}^{(1)}$ term are
	\begin{align}
		c_{r} &= \frac{\sigma}{2} \left( 1 + \frac{\sigma}{6 \beta} \right) \\
		c_{i} &= \omega_H + \frac{\sigma^2}{36 \omega_H}
			+ \frac{\sigma \omega_H}{6 \beta} \left( 1 - \frac{\sigma}{3 \beta} \right).
	\end{align}
It is convenient to define an effective diffusion constant $\delta$ in terms of the divorced mobility rates $\delta_D$ and $\delta_E$ such that
	\begin{equation}
		\label{equ:cgle_delta}
		\delta = \frac{3 \beta \delta_E + \sigma \delta_D}{3 \beta + \sigma}.
	\end{equation}
The form of the combined constant $\delta$ gives clues to the contributions from the two diffusion rates weighted by the reaction rates $\beta$ and $\sigma$. This shows an intuitive relation between migration and biological processes. For example, when reproduction is high for $\beta \gg \sigma$, exchange of habitat dominates due to lack of empty space. On the other hand, when $\beta \ll \sigma$, diffusive migration dominates as aggressive predation leaves the ecosystem mostly unoccupied. Nevertheless, $\delta$ can be set to unity by rescaling $\bm{X}$ which changes the sizes of the overall patterns in the domain without affecting their dynamics (see main text).

Finally, Eq.~\eqref{equ:cgle_two_c} is simplified by rescaling $\mathcal{A}^{(1)} \rightarrow \mathcal{A}^{(1)} / \sqrt{c_{r}}$  and introducing the sole parameter $c =c_{i} / c_{r}$ to give the final form of the CGLE (\ref{CGLE}).
Thus, the remaining parameter $c$ combines the reaction rates from the generic metapopulation model in the following way
	\begin{equation*}
	\label{equ:cgle_c}
		c = \frac{c_{i}}{c_{r}} =
		\frac
		{12\zeta (6\beta - \sigma)(\sigma + \zeta) + \sigma^2 (24\beta - \sigma)}
		{3\sqrt{3} \sigma (6\beta + \sigma)(\sigma + 2\zeta)},
	\end{equation*}
which is the expression of (\ref{c3}).

\begin{thebibliography}{99}
\bibitem{SAnonHB}
It is indeed worth noting that we cannot rule out that spiral annihilation may also occur at low mutation rate after a much longer simulation time than in the proximity of HB. Yet, to the best of our knowledge no spiral annihilation phase has been observed in the studies carried out in the absence of mutations ($\mu=0$), see e.g.~\cite{RMF,RF,Matti,He,lamouroux12,rulands13}.
\end{thebibliography}
\end{document}